\documentclass[fleqn,usenatbib]{mnras}

\usepackage[usenames,dvipsnames]{xcolor}
\usepackage{amsmath}
\usepackage{array}
\usepackage{xfrac}
\usepackage{multirow}

\usepackage[T1]{fontenc}
\usepackage{ae,aecompl}
\hypersetup{draft}

\title[Predicting MaNGA gas gradients]{How well do local relations predict gas-phase metallicity gradients? Results from SDSS-IV MaNGA}

\author[Boardman et al.]{
N.~Boardman$^{1}$\thanks{E-mail: nfb@st-andrews.ac.uk},
G.~Zasowski$^{2}$,
J.~A.~Newman$^{3}$,
S.~F.~Sanchez$^{4}$,
B.~Andrews$^{3}$,\and
J.~K.~Barrera-Ballesteros$^{4}$,
J.~Lian$^{5}$,
R.~Riffel$^{6,7}$,
R.~A.~Riffel$^{7,8}$,
A.~Schaefer$^{9}$,
K.~Bundy$^{10}$\\
$^{1}$School of Physics and Astronomy, University of St Andrews, North Haugh, St Andrews KY16 9SS, UK\\
$^{2}$Department of Physics \& Astronomy, University of Utah, Salt Lake City, UT, 84112, USA\\
$^{3}$Department of Physics \& Astronomy and PITT PACC, University of Pittsburgh, Pittsburgh, PA 15260, USA\\
$^{4}$Universidad Nacional Autónoma de México, Instituto de Astronomıa, A.P. 70-264, 04510, Mexico, D.F., Mexico\\
$^{5}$Max Planck Institute for Astronomy, 69117, Heidelberg, Germany\\
$^{6}$ Departamento de Astronomia, Instituto de F\'\i sica,
Universidade Federal do Rio Grande do Sul, CP 15051, 91501-970, Porto
Alegre, RS, Brazil \\
$^{7}$Laborat\'orio Interinstitucional de e-Astronomia - LIneA, Rua
Gal. Jos\'e Cristino 77, Rio de Janeiro, RJ - 20921-400, Brazil\\
$^{8}$Departamento de F\'isica, CCNE, Universidade Federal de Santa Maria, 97105-900, Santa Maria, RS, Brazil\\
$^{9}$Max-Planck-Institut f\"ur Astrophysik, Karl-Schwarzschild-Str. 1, D-85748 Garching, Germany\\
$^{10}$UCO/Lick Observatory, University of California, Santa Cruz, 1156 High St. Santa Cruz, CA 95064, USA
}

\date{Accepted 2022 May 25. Received 2022 May 25; in original form 2021 August 5}
\pubyear{2021}
\begin{document} 
\label{firstpage}
\pagerange{\pageref{firstpage}--\pageref{lastpage}}
\maketitle

\begin{abstract}

Gas-phase metallicity gradients in galaxies provide important clues to those galaxies' formation histories. Using SDSS-IV MaNGA data, we previously demonstrated that gas metallicity gradients vary systematically and significantly across the galaxy mass--size plane: at stellar masses beyond approximately $10^{10}$ $\mathrm{M_\odot}$, more extended galaxies display steeper gradients (in units of $\mathrm{dex/R_e}$) at a given stellar mass. Here, we set out to develop a physical interpretation of these findings by examining the ability of local $\sim$kpc-scale relations to predict the gradient behaviour along the mass--size plane. We find that local stellar mass surface density, when combined with total stellar mass, is sufficient to reproduce the overall  mass--size trend in a qualitative sense. We further find that we can improve the predictions by correcting for residual trends relating to the recent star formation histories of star-forming regions. However, we find as well that the most extended galaxies  display steeper average gradients than predicted, even after correcting for residual metallicity trends with other local parameters. From these results, we argue that gas-phase metallicity gradients can largely be understood in terms of known local relations, but we also discuss some possible physical causes of discrepant gradients. 

\end{abstract}


\section{Introduction}

Star formation is amongst the most fundamental processes in galaxy evolution, but the physics behind it remain to be fully characterised. As the products of successive stellar generations, gas-phase metallicities in galaxies provide important clues to this end. In turn, gas-phase  gradients provide important insights into how galaxies assemble their mass and structure over time.

A number of recent studies have investigated gas-phase metallicity gradients in galaxies, aided significantly by large integral-field unit (IFU) spectroscopy surveys such as CALIFA \citep{sanchez2012}, SAMI \citep{croom2012}, and MaNGA \citep{bundy2015}. \citet{sanchez2014} report a characteristic gas metallicity gradient of 0.1 dex per disc effective radius in their sample of CALIFA galaxies  \citep[see also][]{sanchez2012b}. Subsequent CALIFA analyses suggest a connection with morphology \citep[e.g.,][]{sm2016}, while \citet{belfiore2017} and \citet{mingozzi2020} report a mass-dependence for gas metallicity gradients in MaNGA galaxies. \citet{carton2018} likewise report a mass-dependence on gradients in their sample of MUSE galaxies, while also finding larger galaxies to display steeper size-scaled gradients on average; a similar size dependence was noted in \citet{boardman2020} for a sample of MaNGA-observed Milky Way analog galaxies. On the other hand, \citet{sm2018} find a tight relationship between galaxies' disc effective radii and the radii at which metallicities decay by 0.1 dex, implying no such size dependence in gas metallicity gradients for their own MUSE sample. \citet{franchetto2021} meanwhile find gas metallicity gradients to be steeper, on average, for galaxies of higher gas fractions at a given stellar mass.

Recently, we showed in \citet[][hereafter B21]{boardman2021} that gas-phase metallicity gradients vary systematically across the galaxy mass-size plane, using a sample of 1679 star-forming MaNGA galaxies. We found a clear pattern beyond stellar masses of around $\mathrm{10^{10} M_\odot}$, wherein more extended galaxies displayed steeper gradients on average (in units of dex per effective radius, hereafter $\mathrm{dex/R_e}$) at a given stellar mass. We further demonstrated that this behaviour cannot be explained as being purely a result of the MaNGA point spread function (PSF), that it is not simply a reflection of known morphology trends, and that it is not dependent on the choice of galaxy size parameter. Thus, we argued this behaviour to be physical in nature. We further argued this result to be consistent with previous findings of a morphology-gradient connection, given the known connection between morphology and size at a given galaxy mass \citep{fl2013}. 

We did not offer a detailed physical explanation in B21 for the observed gradient behaviour. Such an explanation would be timely, particularly in light of the numerous models and simulations available for comparison with observations. Galaxy gas metallicity gradients have been studied in semi-analytic galaxy models \citep[e.g.,][]{yates2021} as well as in full cosmological hydrodynamical simulations \citep[e.g.,][]{tissera2019}. Thus, by better understanding the observed gas metallicity behaviour of galaxies, we may then provide some powerful comparisons to be made with simulations, and thus provide a tool for further understanding how gas-phase abundances evolve over time.


Gas-phase metallicity has been shown to be related to a number of other parameters in galaxies across $\sim$kpc scales. In particular, a strong relation has been repeatedly found between the gas-phase metallicities and stellar mass surface densities ($\Sigma_*$) of star-forming regions \citep[e.g.,][]{ro2012,sanchez2013,bb2016}. This can be viewed as a ``local" version of the well-known mass-metallicity relation \citep[e.g.,][]{lequeux1979,tremonti2004}, and indeed appears sufficient to reproduce that very relation \citep{ro2012}; we refer to this local relation as the ``resolved mass-metallicity relation" (rMZR) over the remainder of this work. The rMZR been argued by \citet{bb2016} to be sufficient to reproduce galaxies' gas metallicity gradients for all but the least massive MaNGA galaxies. An additional dependence between gas metallicity and galaxy stellar mass ($\mathrm{M_*}$) is also apparent at a given value of $\Sigma_*$, particularly at lower masses  \citep[e.g.,][]{bb2016,gao2018,hwang2019}. 

In the majority of star-forming MaNGA galaxies, an anti-correlation can be seen between local star formation rates (SFRs) and gas metallicities in star-forming regions once radial trends in both properties are removed \citep[][]{sm2019} -- a result that has also been obtained in the EAGLE cosmological simulations \citep{sd2021}. This anti-correlation can be understood from mathematical arguments \citep{sa2019} as being directly related to the Fundamental Metallicity Relation \citep[FMR;][]{ellison2008,mannucci2010,ll2010}, in which galaxies with higher SFRs display lower gas metallicities at a given stellar mass.The existence of such a ``local FMR" is not detected in the CALIFA dataset \citep{sanchez2013}, however, and appears to depend on the adopted metallicity calibrator within the MaNGA dataset \citep{teklu2020}. The FMR also appears to be scale-dependent, with a \textit{positive} metallicity-SFR residual trend instead found on $\sim$100 pc scales \citep{wang2021}. 

Various other trends involving gas metallicity have been reported, both at local and global galaxy scales. \citet{hwang2019} find low values of $\mathrm{D_n4000}$\footnote{In this paper, we use D4000 to refer to the original \citet{bruzual1983} index definition and $\mathrm{D_n4000}$ to refer to the later narrow-band definition of \citet{balogh1999}.} to be associated with lower gas metallicities at a given combination of $\Sigma_*$ and $\mathrm{M_*}$, with \citet{sm2020} similarly reporting positive correlations between D4000 and gas metallicities in the majority of their sample MaNGA galaxies; this again has a global equivalent within the mass-metallicity relation, in which low $\mathrm{D_n4000}$ values are associated with lower gas-phase metallicities at a given mass \citep{lian2015}. \citet{bb2018} report gas metallicity to correlate with both gas mass fraction (negative correlation) and local escape velocity (positive correlation), though they note the trends to be weaker than the density-metallicity relation. Overall, these findings indicate a significant connection between the gas metallicity of a galaxy region and the formation history of that region; such a connection, along with its ability to predict the mass--size behaviour of gas metallicity gradients, is the focus of this present work. 

Here, we assess the ability the local relationships to reproduce the observational B21 results and set out to develop a physical interpretation of the observed gradient trends. We present the data to be used in our study in \autoref{sample}. We discuss local trends and present the resulting metallicity predictions in \autoref{models}, and we discuss our results in terms of predicted metallicity gradients in \autoref{results}. We discuss our findings in \autoref{discussion}, and then summarise and conclude in \autoref{conclusion}. We assume the standard $\Lambda$ Cold Dark Matter cosmology throughout this work, and we adopt the following parameters: $\mathrm{H_0} = 71$ km/s/Mpc, $\mathrm{\Omega_M}$ = 0.27, $\mathrm{\Omega_\Lambda}$ = 0.73.

\section{MaNGA galaxy sample and other data.}\label{sample}

\subsection{MaNGA data}\label{data}

We begin with the parent galaxy sample presented in B21, consisting of SDSS-IV MaNGA galaxies with axis ratios (b/a) no lower than 0.6 and with stellar masses  available from the GALEX-SDSS-WISE Legacy catalog \citep[GSWLC;][]{salim2016, salim2018}. We obtain $\mathrm{M_*}$ values from the GSWLC-2X table \citep{salim2018} and convert these from a \citet{chabrier2003} IMF to a Kroupa \citep{kroupa2001, kroupa2003} IMF, as in B21; we perform this conversion by multiplying by a factor of 1.06 \citep{salim2007,elbaz2007, zahid2012,speagle2014}. We use for $\mathrm{R_e}$ the elliptical Petrosian half-light radii obtained from the Nasa-Sloan-Atlas \citep[NSA;][]{blanton2011} catalog. We use elliptical Petrosian b/a values and position angles (PAs) from the NSA catalog. 

The MaNGA galaxies were observed with the BOSS spectrographs \citep{smee2013} on the 2.5~m Sloan telescope at Apache Point Observatory \citep{gunn2006}. The MaNGA IFUs contain 19--127 optical fibres of diameter 2$^{\prime\prime}$ each, in hexagonal configurations; observations with these IFUs employ a three-point dithering pattern to fully sample the field of view \citep{drory2015,law2015}. The observations are reduced with the MaNGA Data Reduction Pipeline \citep[DRP;][]{law2016,yan2016a}, which yields $0.5^{\prime\prime} \times 0.5^{\prime\prime}$ spaxel datacubes with a median PSF full-width at half-maximum (FWHM) of approximately 2.5$^{\prime\prime}$ \citep{law2016}. The individual reduced spectra have a spectral resolution of $R \simeq 2000$ \citep{smee2013}, with a wavelength range of 3600--10000\AA . The MaNGA galaxy sample, consisting of roughly 10,000 galaxies in all, was selected to have a roughly flat distribution of log-mass and a redshift range of approximately 0.01 to 0.15 \citep{yan2016b,wake2017}. A number of spaxel-based quantities relating to stellar and gaseous features are computed by the MaNGA Data Analysis Pipeline \citep[DAP;][]{belfiore2019,westfall2019}, and are available via the Marvin\footnote{\url{https://www.sdss.org/dr16/manga/marvin/}} interface \citep{cherinka2019}. MaNGA results first became publicly available in SDSS Data Release 15 \citep[DR15;][]{aguado2019}, with further data -- along with DAP results -- released in SDSS DR16 \citep{ahumada2020} and DR17 \citep{sdssdr17}.

We make use here of several spaxel properties derived by the Pipe3D analysis pipeline \citep{sanchez2016,sanchez2016b,sanchez2018} and by the MaNGA DAP, both to assess local relations involving gas metallicity and to construct predictions of gas metallicity using those relations. We employ the MaNGA Product Launch 10 (MPL-10) versions of these pipelines. In general, we choose to focus on properties that relate to the star formation history (SFH) of a region and/or to the gas contents of that region. The Pipe3D properties have been derived in bins with target signal-to-noise (S/N) values of 50, while in the case of the DAP we use unbinned values in all cases. We correct emission line fluxes for reddening by assuming an intrinsic Balmer decrement of 2.86 (valid for case B recombination, with $\mathrm{n_e = 100 \ cm^{-3}}$ and $\mathrm{T_e = 10000 \ K}$) along with a \cite{calzetti2000} attenuation curve; we assume $R_V$ = 3.1 when performing this correction, following e.g., \citet{greener2020}. We use the resulting reddening-corrected flux measurements throughout our analysis. 

We obtain the following values from Pipe3D:

\begin{itemize}
    \item \textbf{Stellar surface mass density} ($\Sigma_*$): this is computed from Pipe3D fits to the stellar continuum, which include a dust correction. As in \citet{bb2016}, we multiply the observed densities by b/a to correct for inclination; we use these corrected values for the remainder of this article. We also convert from a \citet{salpeter1955} IMF to a Kroupa IMF by multiplying by a factor of 0.62 \citep{salim2007,elbaz2007, zahid2012,speagle2014}, for consistency with the total stellar masses employed in this work. Finally, we multiply the densities through by each galaxy's Pipe3D dezonification map (as described in Section 3.4.4 of \citealt{sanchez2016b}), to account for the effects of spatial binning on galaxy's derived density maps. 
     \item \textbf{D4000 index}: we take this parameter directly from Pipe3D, with no corrections or adjustments performed. The D4000 index is calculated as the ratio between the average flux densities at 3750-3950 \AA\ and 4050-4250 \AA\, following the definition of \citet{bruzual1983}. D4000 positively correlates with the light-weighted stellar age of a galaxy region (see, for instance, Figure 12 of \citet{sanchez2016b}) and also functions as an estimator of a region's specific star formation rate, though on longer timescales than is measured by $H\alpha$ emission flux. We stress however that D4000 is \textit{not} purely an indicator of age or star formation, as increased metal absorption from metal-rich stars will also lead to higher D4000 values.
    \item \textbf{Light-weighted stellar age} ($t_{LW}$): this is calculated as the light-weighted combination of values from simple stellar population (SSP) template fits to observed spectra.
\end{itemize}

We obtain the following parameters from the DAP:

\begin{itemize}
    \item\textbf{Star formation rate surface density} ($\Sigma_{SFR}$): we calculate this from spaxels' non-parametic $H\alpha$ emission fluxes, employing Equation 20 of  \citet[][which assumes a Kroupa IMF]{kennicutt2009}. We multiply our obtained values by b/a to correct for inclination, as with $\Sigma_*$, and we use these corrected values over the remainder of this article. In addition to $\Sigma_{SFR}$ itself, we also make use of the local specific star formation rate ($\mathrm{sSFR_{local}}$), defined as $\Sigma_{SFR} / \Sigma_{*}$, in our analysis. 
    \item \textbf{H$\mathbf{\alpha}$ equivalent width}, hereafter EW(H$\alpha$): we obtain this directly from the DAP, employing the non-parametric measurements. This is essentially a proxy for $\mathrm{sSFR_{local}}$; however, $\mathrm{sSFR_{local}}$ is vulnerable to error propagation from the calculation of $\Sigma_*$ \citep[e.g.][]{hwang2019} and from the dust reddening, so we choose to consider EW(H$\alpha$) in addition to $\mathrm{sSFR_{local}}$.
    \item \textbf{Optical extinction} $A_V$, measured from the Balmer decrement as part of our reddening correction. This parameter serves as a reasonable proxy for gas mass within a region \citep[e.g.][]{bb2018}.
    \item \textbf{Observed gas-phase metallicity}, $\rm 12 + log(O/H)_{obs}$. We calculate metallicity using the O3N2 calibrator of \citet[][hereafter M13]{marino2013} along with the R calibration described in Equations 4 and 5 of \citet[][hereafter PG16]{pilyugin2016}. We will focus on results from the M13 calibrator in main text of this article, but we present results from the PG16 calibrator in Appendix B.
\end{itemize}

In addition to the parameters listed above, we also consider the ratios $A_V / M_*$ and $\Sigma_{SFR}/A_V $, which serve as proxies for the gas-to-stellar mass ratio and the star formation efficiency (SFE) respectively \citet[e.g.][and references therein]{bb2018}.

\subsection{Identification of star-forming regions}

Gas metallicity emission line calibrators are generally only valid in regions dominated by star formation \citep[though, see][]{kumari2019}; thus, for the remainder of our analysis, we restrict to star-forming spaxels with well-measured emission lines. We select star-forming spaxels by requiring that their emission ratios fall within the \citet{kauffmann2003} star-forming region on the BPT-NII diagram \citep{bpt}, and by requiring a minimum EW(H$\alpha$) of 10\AA. We also removed spaxels with measured Balmer ratios of less than 2.86 before dust-correction, considering them unreliable for our purposes. 

To ensure reliable metallicity measurements from both the M13 and PG16 calibrators, we restrict to spaxels for which the following emission features are detected with S/N $> 3$: H$\alpha$, H$\beta$, [OIII]$_{5008}$,[NII]$_{6585}$, [OII]$_{3737, 3729}$. We further restrict to spaxels located between 0.5 $\mathrm{R_e}$ and 2 $\mathrm{R_e}$; this is because our metallicity gradients are calculated over this range, as described in \autoref{gradcalc}. We remove spaxels with unreliable D4000 measurements (for which we adopt the loose requirement that 0 < D4000 < 2), and we remove spaxels with M13-derived metallicities beyond the original fitted range of the M13 O3N2 calibrator ($8.17 \leq 12+\log(\mathrm{O/H})_{obs} \leq 8.77$).  Finally, we only consider star-forming spaxels within galaxies with at least 20 such spaxels; this is because the angular area of the MaNGA PSF is approximately 20 spaxels, as for instance pointed out by \citet{hwang2019}. We obtain from these restrictions a sample of 871346 spaxels overall.

\subsection{Gas metallicity gradient calculation}\label{gradcalc}

We calculate our galaxies' radial gas-phase metallicity gradients, $\rm \nabla [O/H]_{obs}$, in units of $\mathrm{dex/R_e}$, from spaxels' gas-phase metallicities at radii between 0.5 $\mathrm{R_e}$ and 2 $\mathrm{R_e}$.  We perform this calculation using a least-absolute-deviation fit, for all galaxies with at least twenty available star-forming spaxels. Our chosen radius range minimises the impact of PSF effects on calculated gradients, while also serving to avoid significant breaks from linear fits \citep{sanchez2014,sm2016,belfiore2017}.

We estimate errors using a bootstrapping analysis: we randomly resample the residuals from the best-fit line 100 times apiece and re-fit the metallicity gradient, before taking as the error the dispersion of the re-fit gradients. We calculate the dispersion using the ``ROBUST\_SIGMA" IDL procedure \footnote{https://idlastro.gsfc.nasa.gov/ftp/pro/robust/robust\_sigma.pro}, which we also use for all subsequent dispersion calculations; this procedure calculates a dispersion estimate that is resistant to outliers and is equivalent to a standard deviation calculation in the case of an outlier-free distribution. We discount any galaxy for which the gradient error is greater than 0.1 $\mathrm{dex/R_e}$. As discussed further in \autoref{modelssum}, this results in a final sample of 2123 galaxies containing 861134 star-forming spaxels.

\begin{figure*}
\begin{center}
	\includegraphics[trim = 0.5cm 16.5cm 1cm 7cm,scale=1.2,clip]{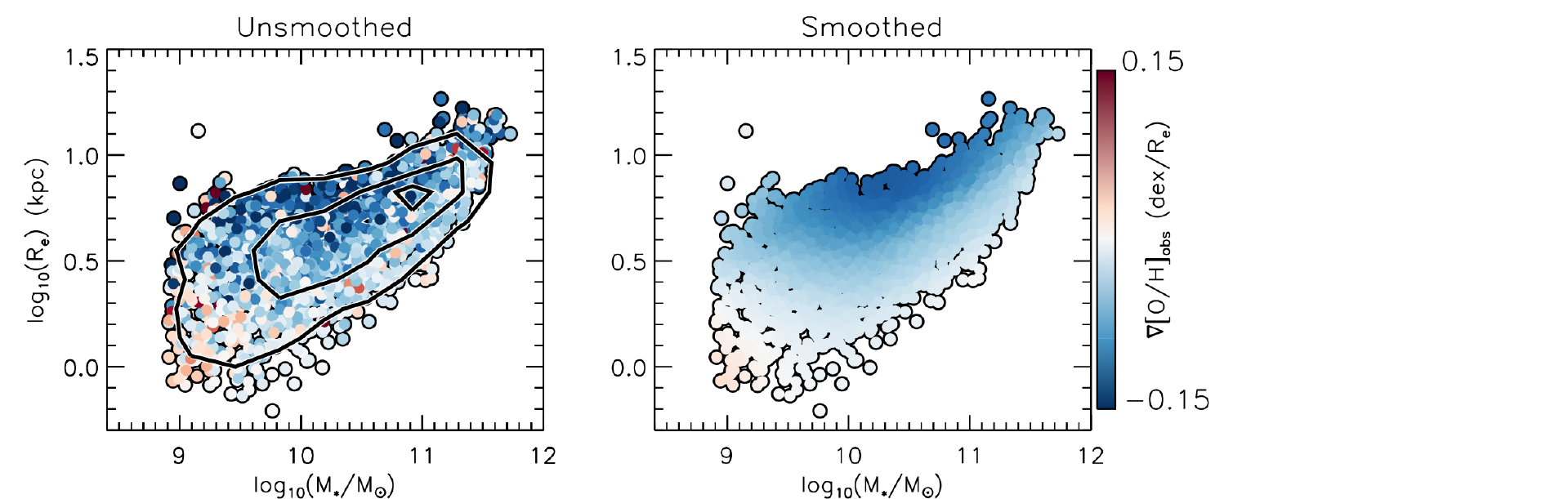}
	\caption{Effective radius plotted against galaxy stellar mass, with data points coloured by the observed gas metallicity gradients before (left) and after (right) applying LOESS smoothing. The contours on the top panel encompass $\sim$90\%, $\sim$50\% and $\sim$10\% of sample galaxies.}
	\label{fig1}
	\end{center}
\end{figure*}

In \autoref{fig1}, we present the $\rm \nabla [O/H]_{obs}$ values of our galaxies as a function of $\mathrm{R_e}$ and $\mathrm{M_*}$; this is the same empirical trend that was explored in B21, but with our own gradient calculations instead of gradients obtained from the Pipe3d summary table \citep{sanchez2018}. We show the raw gradients in the left panel, and in the right panel we show the results of applying two-dimensional locally-weighted regression smoothing \citep[LOESS;][]{cleveland1988} as implemented in IDL\footnote{available from \url{http://www-astro.physics.ox.ac.uk/~mxc/software/}}. We compute the LOESS-smoothed value for each data point using the closest 20\% of data points with the \textit{rescale} keyword applied, with errors computed from the scatter in neighbouring points for the purpose of the calculation. As in B21, we see that galaxies' gas metallicity gradients relate to \textit{both} stellar mass and size: the gradients steepen with size at a given stellar mass, pariticularly for stellar masses above approximately $10^{10}$ $\mathrm{M_\odot}$, and also steepen with mass at low masses. 

We are implicitly assuming every spaxel to be resolved for the purpose of our gradient calculation. Given the MaNGA PSF, such an assumption does not actually hold across neighbouring spaxels. We have verified however that we obtain similar gradient results if we azimuthally average metallicities over radial bins before calculating gradients. As such, the precise calculation method is not a critical factor in our analysis.

\section{Predicting gas metallicity with local relations}\label{models}

We now employ a number of local $\sim$kpc-scale trends to predict spaxel gas metallicities, in order to test the ability of the trends to predict galaxies' gas metallicity gradients. We refer to predicted gas metallicities as ``model" metallicities for the remainder of this article. 


In general, our methodology in this section is intended to \textit{let the data speak for itself}: we consider the observed gas metallicity as a function of various combinations of parameters, without any fitting of functional forms. Our method of constructing these models is described over the remainder of this section. We also assign ``errors" to each individual model value for the purpose of estimating errors in predicted metallicity gradients; these are determined based on the metallicity scatter in the local relations, as discussed further over this section.

\subsection{Base models}\label{modelsbase}

\begin{figure*}
\begin{center}
	\includegraphics[trim = 1.5cm 11.5cm 1cm 13cm,scale=1]{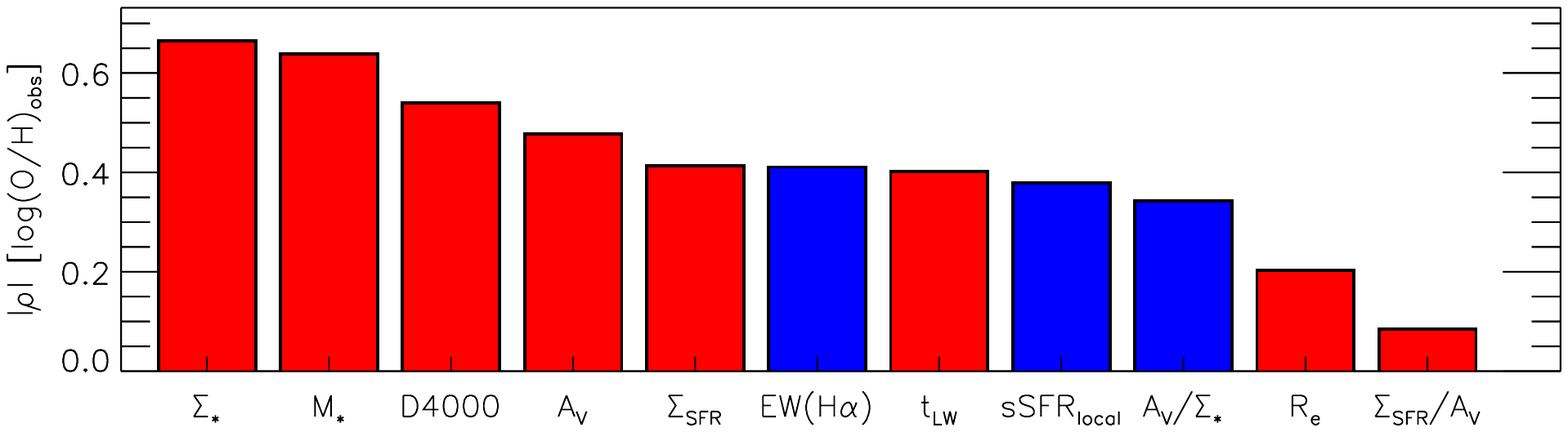} 
	\caption{Absolute values of the Spearman correlation coefficient between the gas-phase metallicity and various other parameters. Red bars indicate positive coefficients, and blue bars negative coefficients. We obtain $p \ll 0.01$ in all cases.}
	\label{obscoeff}
	\end{center}
\end{figure*}

In \autoref{obscoeff}, we plot the Spearman correlation coefficient $\rho$ between observed spaxel metallicities and the corresponding galaxy $M_*$ and $R_e$ values along with all other considered spaxel properties. It is $\Sigma_*$ and then $M_*$ that correlate most strongly with the observed metallicity, out of all observational parameters that we consider. All $\rho$ values in this figure have corresponding p-values of $p \ll 0.01$, as do all subsequently-presented $\rho$ values unless otherwise stated. We note that even low ($\rho < \sim 0.1)$ correlation coefficients are associated with very small p-values in our calculations, due to the large number of spaxels in our sample.

Given the behaviour of our selected dataset, $\Sigma_*$ and $M_*$ make for a natural starting point for our model metallicities. \citet{bb2016} have previously argued the rMZR alone to be sufficient for reproducing galaxies' gas-metallicity gradients for all but the least massive galaxies. However, a significant mass dependence on the rMZR has been reported before in the literature, including by \citet{bb2016} themselves; this suggests that the three-way dependence between metallicity, mass and density is more fundamental than the rMZR alone \citep{gao2018}. More accurate metallicity predictions can therefore be obtained by also taking mass into account, as for instance done by \citet{hwang2019}. 

We thus chose to construct an initial set of model metallicities using a combination of $\Sigma_*$ and $M_*$. Specifically, we compute the mean metallicity in small bins of $\Sigma_*$ and $M_*$, before applying weak LOESS smoothing \footnote{in which we consider the nearest 2\% of bins for each bin, with the \textit{rescale} keyword applied} over the mean values.  For a given spaxel, we then ``predict" the gas metallicity as being equal to the mean metallicity of the associated smoothed mass-density bin. We require a minimum of 10 star-forming spaxels in a given bin in mass-density space, discarding spaxels that do not fall into a bin meeting this criterion. We assign errors as the standard deviations of metallicities within a given bin, with no smoothing applied. 

\begin{figure}
\begin{center}
	\includegraphics[trim = 2cm 18cm 1cm 5.5cm,scale=0.95,clip]{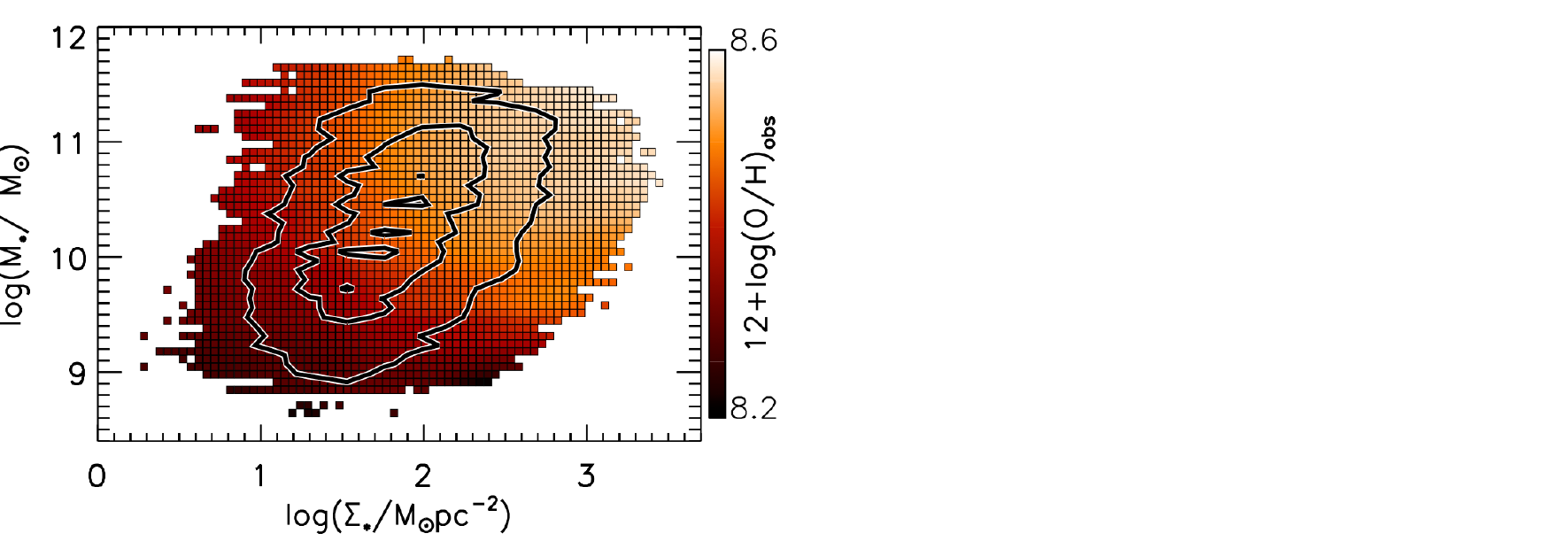} 
	\caption{Mean observed gas metallicity, as a function of $M_*$ and $\Sigma_*$, with weak smoothing applied. The contours encompass $\sim$10\%, $\sim$50\% and $\sim$90\% of sample galaxy spaxels. The plotted relationship is used in making our base models along with all subsequent model sets.}
	\label{rhomfig}
	\end{center}
\end{figure}

We demonstrate our process in \autoref{rhomfig}. These model metallicities are included in the calculation of all others presented in this section. Thus, we refer to them as ``base models" ($\mathrm{12 + \log(O/H)_{base}}$) for the remainder of this work.

\subsection{Extensions to base models}\label{modelsext}

We now consider possible extensions to the base models, based on residual dependencies with other parameters. We explore residual trends in detail here, before summarising in \autoref{modelssum}. 

\begin{figure*}
\begin{center}
	\includegraphics[trim = 1.5cm 11.5cm 1cm 12cm,scale=1]{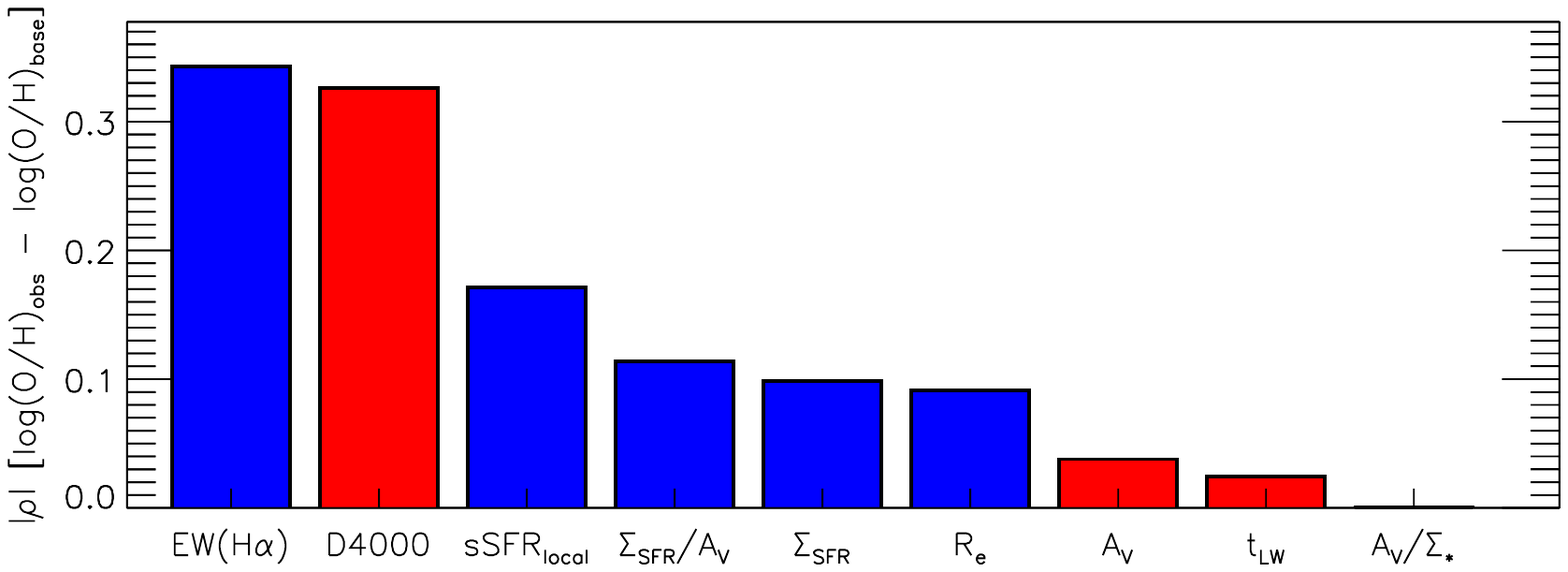} 
	\caption{Absolute values of the Spearman correlation coefficient between the gas-phase metallicity residuals and various other parameters. Red bars indicate positive coefficients, and blue bars negative coefficients. We obtain $p = 0.82$ for $A_V/\Sigma_*$ and $p \ll 0.01$ otherwise.}
	\label{basecoeff}
	\end{center}
\end{figure*}

As a first step, we plot in \autoref{basecoeff} the Spearman correlation coefficient between the metallicity residuals ($\log(O/H)_{obs} - \log(O/H)_{base}$) and all other considered parameters besides $M_*$ and $\Sigma_*$. Similarly to \citet{hwang2019}, we find strong residual metallicity correlations with EW(H$\alpha$) and $D4000$ along with a weaker (though still significant) correlation with $sSFR_{local}$, with dust reddening uncertainties being a likely reason as to why the latter correlation is weaker. We also find very little small residual correlations with $A_V$, $t_{LW}$ or $A_V/\Sigma_*$, with the latter yielding a p-value of $p = 0.82$, in spite of the relatively large $\rho$ values between these parameters and the gas metallicity itself.

The gas mass, we note, is typically a key ingredient in chemical evolution modelling recipes \citep[e.g.][and references therein]{bb2018}, with local SFRs treated as an observable consequence of the gas content. $A_V$, as described in \autoref{data}, serves as a reasonable proxy for gas mass in a star-forming regions. Thus, the apparent lack of a strong residual metallicity dependence in our base models on $A_V$ or $A_V/\Sigma_*$ -- in contrast with $sSFR_{local}$ and associated parameters -- is worth comment. Firstly, we note that the gas fraction and $\Sigma_*$ are quite tightly correlated \citep[e.g.][]{bb2018}, meaning that much of the information available from $A_V$ will already be encoded into our base models. Furthermore, it should be noted that the ratio between gas mass and $A_V$ is not constant in practice, as direct gas measurements show the ratio to increase with EW(H$\alpha$) \citep{bb2020}. Finally, we note somewhat stronger residual trends involving $A_V$ when the alternative \citet{pilyugin2016} R2 calibrator is employed (\autoref{p16appendix}), though we still find stronger residual trends with $sSFR_{local}$ and related parameters. As such, the lack of a strong residual metallicity trend involving $A_V$ for the M13 calibrator should be interpreted with caution.

\begin{figure}
\begin{center}
	\includegraphics[trim = 3.7cm 2.5cm 0cm 13cm,scale=0.9,clip]{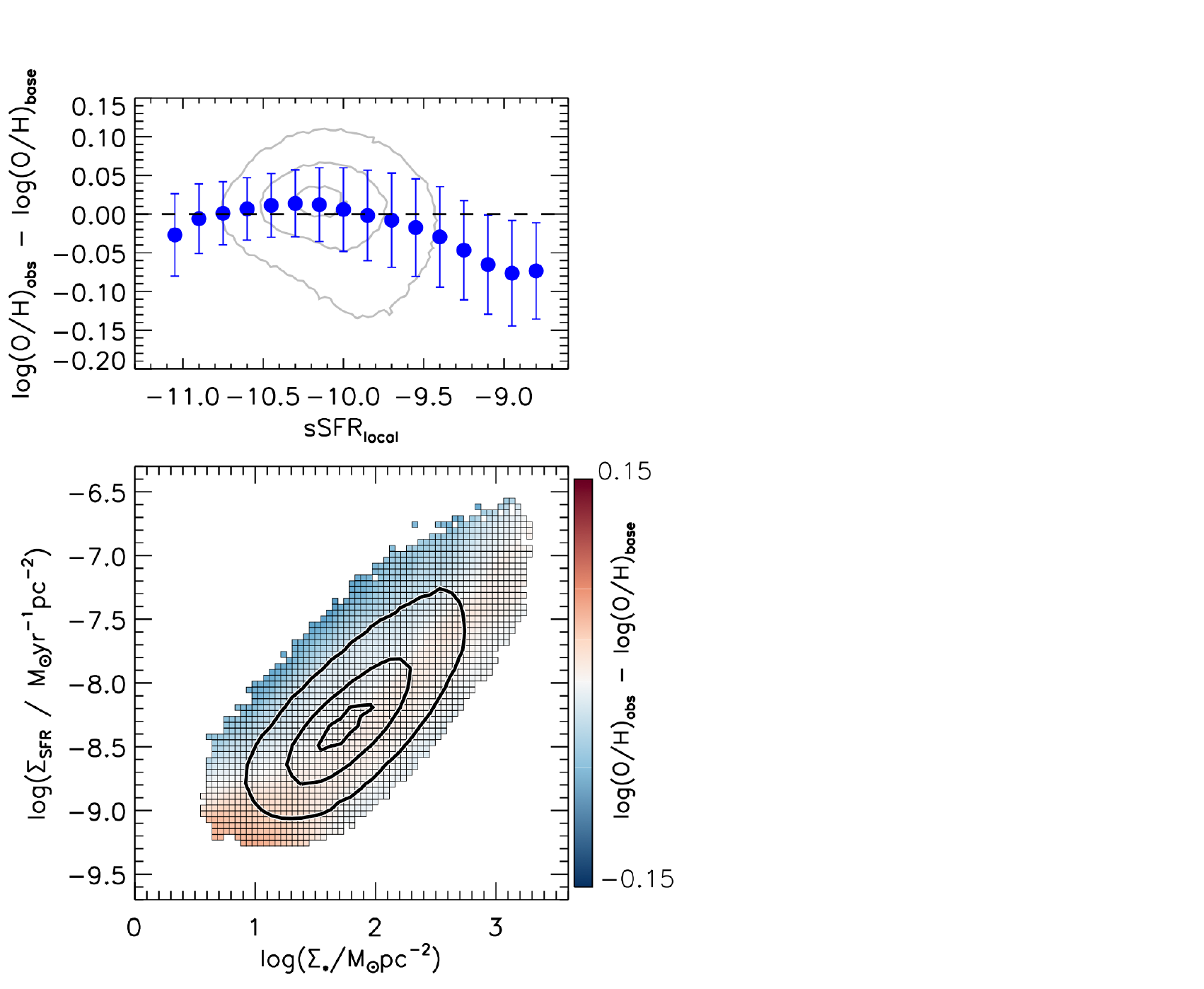} 
	\caption{Mean residuals between observed and base model metallicities, as a function of $sSFR_{local}$ (top) and as a combined function of $\Sigma_*$ and $\Sigma_{SFR}$ with weak smoothing applied (bottom). Contours encompass $\sim$10\%, $\sim$50\% and $\sim$90\% of sample galaxy spaxels. In the top window, data points show medians and dispersions in bins of $sSFR_{local}$ that each contain a minimum of 100 spaxels.}
	\label{rhomsfrres}
	\end{center}
\end{figure}

In \autoref{rhomsfrres}, we plot the residuals between the base models and the observed metallicities, both as a function of $sSFR_{local}$ alone and as a combined function of $\Sigma_{SFR}$ and $\Sigma_*$. For $sSFR_{local}$, we consider bins with a minimum of 100 spaxels; for the $\Sigma_{SFR}-\Sigma_*$ case we show the mean offset in small bins with weak LOESS smoothing applied, for bins containing at least ten spaxels. Our $\Sigma_{SFR}-\Sigma_*$ projection is motivated largely by past investigations of the local FMR, which consider metallicity as a combined function of $\Sigma_*$ and local SFR without considering the corresponding galaxy mass \citep[e.g.][]{teklu2020}. We see that higher $sSFR_{local}$ values are associated on average with lower metallicities relative to what is predicted, with a  mild turnover at the lowest $sSFR_{local}$ values, which corresponds to the appearance of a local FMR in the $\Sigma_{SFR}-\Sigma_*$ parameter space. We therefore find that the local FMR exists in MaNGA data for the O3N2 calibrator, as previously reported by \citet{teklu2020}. Importantly, however, we find that the local FMR is \textit{not} purely a projection of the $M_*$-$\Sigma_*$-$O/H$ relation, and that a residual SFR dependence at a given density remains even when $M_*$ is considered. It is clear, however, that only a minority of spaxels are strongly affected by this particular trend.

\begin{figure}
\begin{center}
	\includegraphics[trim = 1.5cm 2cm 1cm 7cm,scale=0.65]{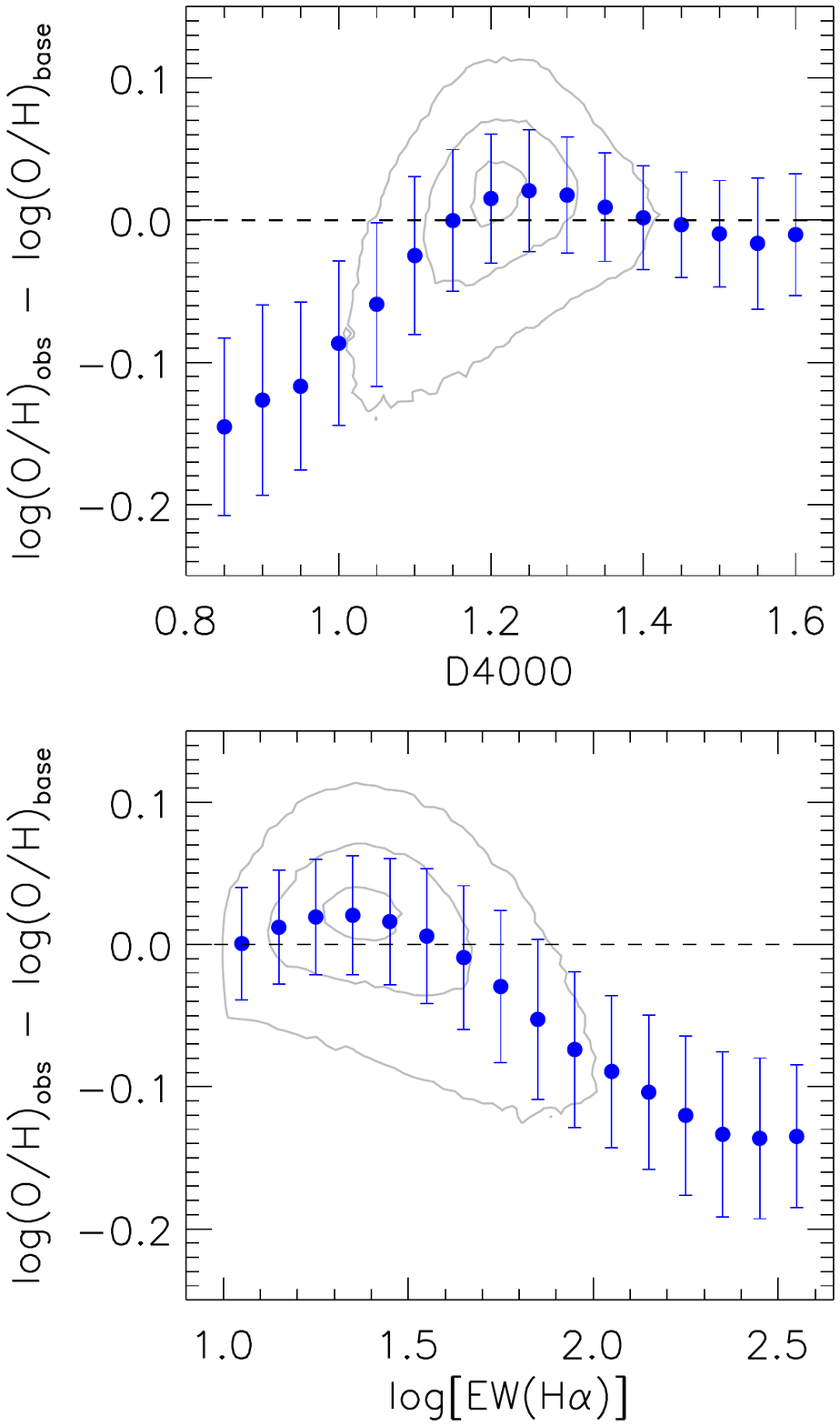} 
	\caption{Residuals between observed and base model metallicities as a function of D4000 (top) and EW(H$\alpha$) (bottom), with data points showing the median and dispersion within a series of bins. Each bin contains at least 100 spaxels. The contours encompass $\sim$10\%, $\sim$50\% and $\sim$90\% of sample galaxy spaxels.}
	\label{d4000res}
	\end{center}
\end{figure}

Next, we consider the base model metallicity residuals in terms of D4000 and EW(H$\alpha$). In \autoref{d4000res}, we plot the residuals as a function of D4000 and as a function of EW(H$\alpha$). As expected, these two residual trends are tighter than that seen for $sSFR_{local}$, making them more suitable for potential extensions to our base models. 

\begin{figure}
\begin{center}
	\includegraphics[trim = 0.9cm 11cm 1cm 10cm,scale=0.8,clip]{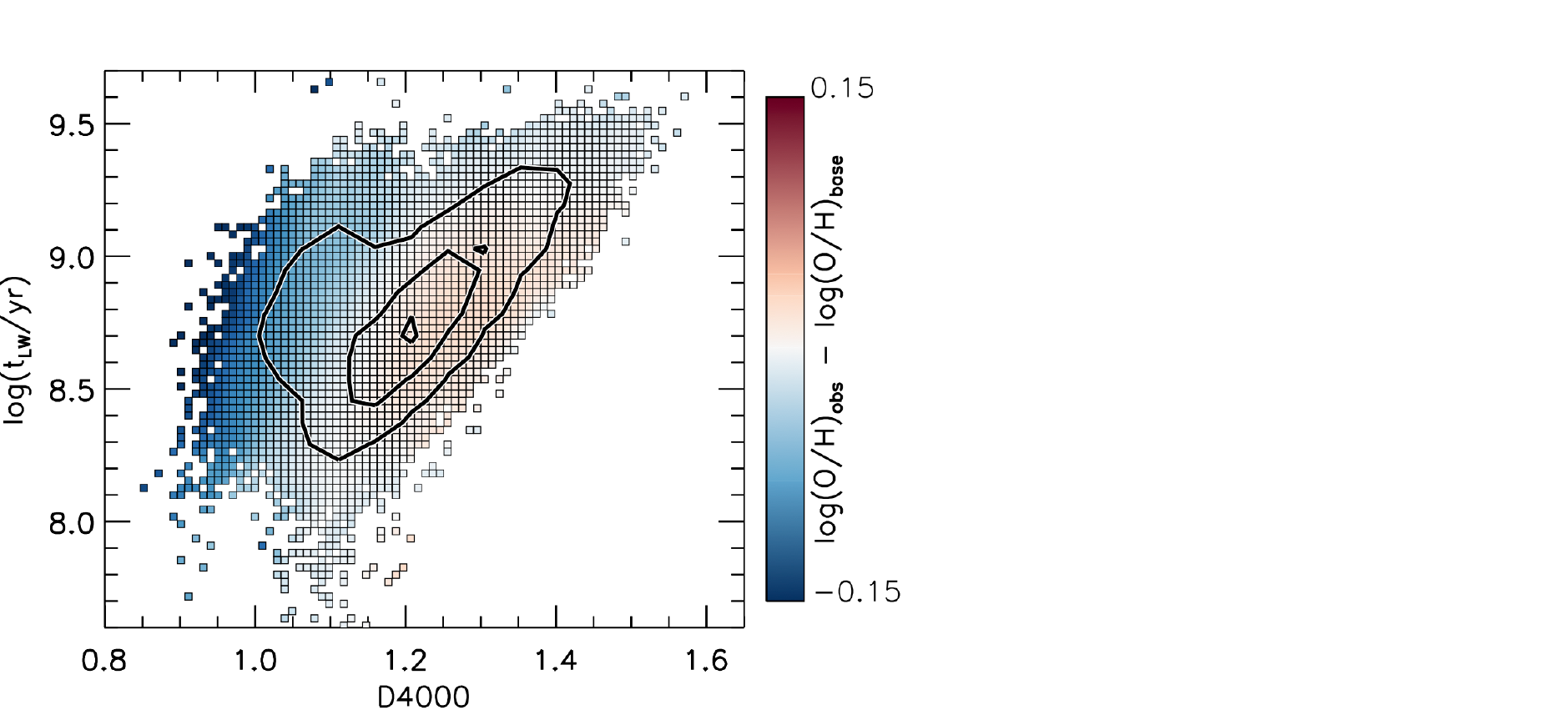} 
	\caption{Mean residuals between observed and base model metallicities, as a function of D4000 and $t_{LW}$ with weak smoothing applied. The contours encompass $\sim$10\%, $\sim$50\% and $\sim$90\% of sample galaxy spaxels. At low D4000 values, a residual $t_{LW}$ trend is apparent.}
	\label{dagelres}
	\end{center}
\end{figure}

\begin{figure}
\begin{center}
	\includegraphics[trim = 0.6cm 11cm 1cm 9.5cm,scale=0.77,clip]{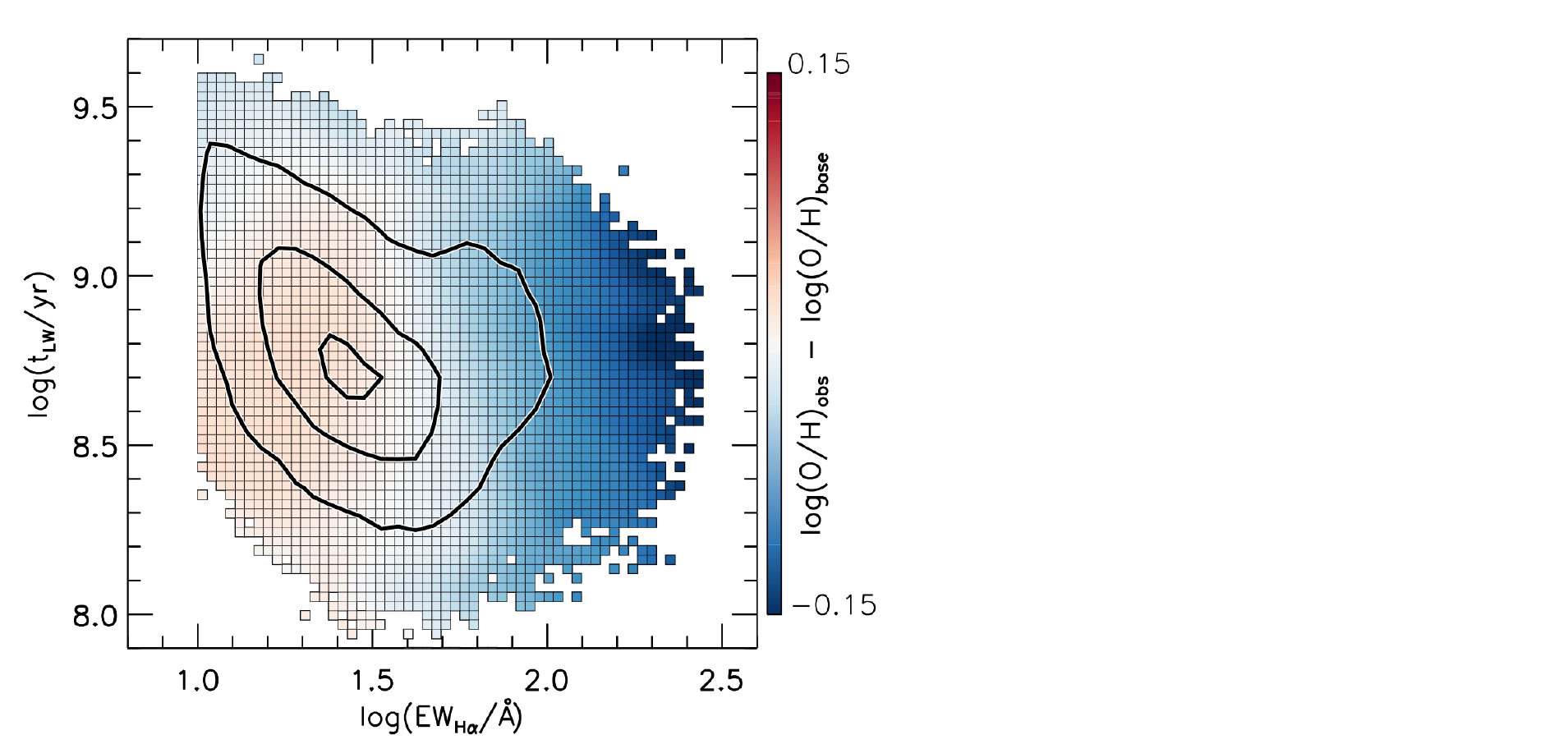} 
	\caption{Mean residuals between observed and base model metallicities, as a function of EW(H$\alpha$) and $t_{LW}$ with weak smoothing applied. The contours encompass $\sim$10\%, $\sim$50\% and $\sim$90\% of sample galaxy spaxels. We find only a mild residual $t_{LW}$ trend for any given value of EW(H$\alpha$).}
	\label{haewagelres}
	\end{center}
\end{figure}

As mentioned previously, we find very little residual dependence on $t_{LW}$; this is in spite of the significant correlation between $t_{LW}$ with the metallicity itself. To investigate $t_{LW}$ further, we plot in \autoref{dagelres} the base model metallicity residuals as a combined function of D4000 and $t_{LW}$. We find a clear two-dimensional trend in the residuals: at low values of D4000, older ages are associated with lower-than-predicted metallicities. If we instead consider the residuals as a combined function of $EW(H\alpha)$, then we detect far less of a residual metallicity dependence on $t_{LW}$ at any value $EW(H\alpha)$; this is shown in \autoref{haewagelres}.

To summarise this subsection so far, we have detected strong residual metallicity trends with EW(H$\alpha$) and D4000 along with a weaker trend with $sSFR_{local}$, in complete agreement with past works. We have further noted a residual metallicity trend with $t_{LW}$ in low-D4000 regions, such that older ages are associated with lower-than-expected observed metallicities. As such, the opportunity exists to construct additional metallicity models with a reduced residual scatter, by incorporating additional parameters besides $\Sigma_*$ and $M_*$.

We experimented with a number of base model extensions by correcting for the base model residuals as a function of one or two parameters. Given the results presented in this section, we focussed on EW(H$\alpha$), D4000 and $t_{LW}$ as potential parameters for this purpose. The models we experimented with are summarised thus:

\begin{itemize}
     \item ``D4000" models. We calculate the median residual between the observed and base model metallicities in bins of D4000, as shown in the top panel of \autoref{d4000res}. We interpolate from bin centres to individual spaxels' D4000 values using linear interpolation. We add these residuals as correction terms to our base model metallicities. Spaxels beyond the the bracketing bins' midpoints do not have model metallicities assigned.
    \item ``EW$(H\alpha)$" models. We construct these in the same manner as for the D4000 models, but by using the EW$(H\alpha)$ bins shown in the bottom panel of \autoref{d4000res}.  
    \item ``SFH" models. We construct these by performing an additive correction to the base models, with the correction equal to the associated D4000-$t_{LW}$ bin residual presented in the bottom panel of \autoref{dagelres}. We require at least ten spaxels per bin, and do not assign model metallicities to spaxels outside of considered bins. These models track recent phases of a region's star formation history (SFH) via D4000 while also taking older phases of the history into account via $t_{LW}$; thus, we refer to these as ``SFH models"
    \item ``SFH-H$\alpha$" models. We construct these in the same manner as the SFH models, except that we instead use the residuals in the EW(H$\alpha$)-$t_{LW}$ bins presented in \autoref{haewagelres}.
\end{itemize}

We assessed the relative merits of these models, along with those of the original base models, using three metrics: the dispersion of residuals, the $\chi^2$ value, and the Bayesian information criterion (BIC). We calculate $\chi^2$ using the measurement error alone; these errors do not take the intrinsic scatter from our chosen calibrator into account, and so result in reduced $\chi^2$ values significantly above unity. We calculate the BIC as $BIC = \chi^2 + k \ln{n}$, where k is the number of estimated model parameters and n the number of spaxels in our sample. Since our models are non-parametric, we take as $k$ the number of parameter bins involved in constructing a given model set (which includes the $M_*-\Sigma_*$ bins used for the base models in all cases); this results in the SFH and SFH-H$\alpha$ models having significantly larger $k$ values than the other models. We also restrict to spaxels with assigned values for \textit{all} model metallicities considered; this results in a sample size $n$ of 847025, which in turn yields $\ln{n} = 13.65$.

\begin{table*}
\begin{center}
\begin{tabular}{c|c|c|c|c|c}
Model set & Model parameters & Residual dispersion & $\chi^2$ & k & BIC \\
\hline
\hline
\textbf{Base models} & $\mathbf{M_*}$\textbf{,} $\mathbf{\Sigma_*}$ & \textbf{0.0523 dex} & $\mathbf{1.14 \times 10^7}$ & \textbf{2061} & $\mathbf{1.14 \times 10^7}$ \\
D4000 models &$M_*$, $\Sigma_*$, D4000 & 0.0465 dex & $5.85 \times 10^6$ & 2077 & $5.88 \times 10^6$ \\
$H\alpha$ models & $M_*$, $\Sigma_*$, EW(H$\alpha$) & 0.0455 dex & $5.49 \times 10^6$ & 2077 & $5.52 \times 10^6$ \\
SFH-H$\alpha$ models & $M_*$, $\Sigma_*$, EW(H$\alpha$), $t_{LW}$ & 0.0452 dex & $5.33 \times 10^6$ & 4515 & $5.39 \times 10^6$ \\
\textbf{SFH models} & $\mathbf{M_*}$\textbf{,} $\mathbf{\Sigma_*}$\textbf{, D4000,} $\mathbf{t_{LW}}$ & \textbf{0.0453 dex} & $\mathbf{5.10 \times 10^6}$  &\textbf{4267} &$\mathbf{5.16 \times 10^6}$ \\
\end{tabular}
\end{center}
\caption{Summary of considered models' details and performance. We find that the SFH models yield the lowest $\chi^2$ and BIC values. Thus, we employ the SFH models over the remainder of this paper, along with the base models for the sake of comparison.}
\label{testtable}
\end{table*}

We summarise the results of our assessment in \autoref{testtable}. Compared to the base models, we find all other models to yield lower residual dispersions. We also find all other models to yield much lower BICs than the base models, indicating that the reduction in dispersion is statistically significant. Out of all models, the SFH models have the lowest BIC as well as the lowest $\chi^2$; thus, we will study the SFH models over the remainder of this article, along with the base models for the sake of comparison.

\subsection{Models summary}\label{modelssum}

To summarise, we have developed in this section a number of sets of model gas metallicities, in which metallicities are predicted using local relations within galaxies. The two model sets to be explored over the remainder of the article are as follows:

\begin{itemize}
    \item $\mathrm{12+\log(O/H)_{base}}$: ``base" gas metallicity models, predicted from spaxel $\Sigma_*$ values and galaxy $M_*$ values.
    \item $\mathrm{12+\log(O/H)_{SFH}}$: ``SFH" models predicted from $\Sigma_*$, $M_*$, D4000 and $t_{LW}$.
\end{itemize}

\begin{figure}
\begin{center}
	\includegraphics[trim = 1.5cm 2cm 1cm 12cm,scale=0.95]{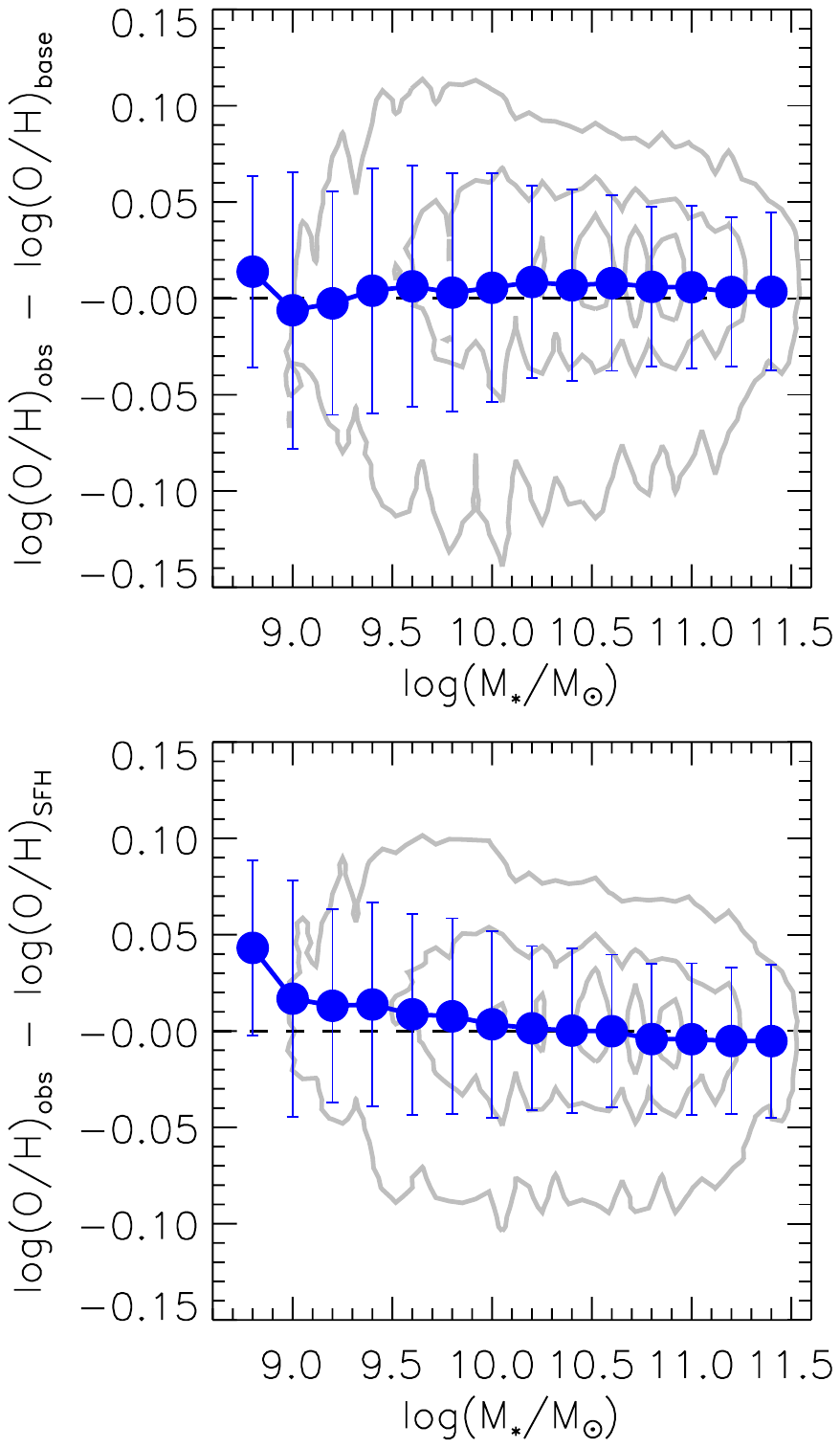} 
	\caption{Gas metallicity residuals plotted against stellar mass for the base models (top) and SFH models (bottom). The contours encompass $\sim$10\%, $\sim$50\% and $\sim$90\% of sample galaxy spaxels.}
	\label{spaxoffsets_fig3}
	\end{center}
\end{figure}

We present in \autoref{spaxoffsets_fig3} the metallicity residuals for both of these model sets as functions of galaxies' stellar mass. The base models by construction display little residual dependence with $\mathrm{M_*}$, but we note a slight residual trend with stellar mass for SFH models. However, the mean SFH model residuals at a given mass remain small compared to the total scatter, for all but the very lowest masses. Motivated by this, we remove from our sample the 15 galaxies with masses below $10^{8.9}M_\odot$, and we do not perform any additional corrections to our models.
    
For our final spaxel sample, we require that spaxels have assigned values for the base and SFH models. After restricting to these samples, we remove from our sample any galaxy that no longer has at least 20 sample star-forming spaxels associated with it. We thus arrive at a final sample of 2123 galaxies containing 861134 spaxels. Our base models yield a median residual of 0.005 dex, while the SFH models yield a median residual of 0.001 dex. We obtain residual  dispersions of 0.052 dex, and 0.045 dex for the base models and SFH models respectively.

As a final note, we found during tests that the precise order of model parameters (that is, which parameters are used in the base models vs which parameters are then corrected on) has only a mild effect on the final SFH model metallicity values. We briefly explore this point in \autoref{ordertest_appendix}, to which we refer the interested reader.

\section{Results}\label{results}

\subsection{Gas metallicity gradient predictions}\label{gradpredictions}

We now test the ability of our model metallicities to reproduce the behaviour of observed gradients across the mass-size plane. For both the base models and SFH models, we calculate gradients in the same manner as for the observations (\autoref{gradcalc}): we perform a least absolute deviation to the predicted metallicities of a galaxy's star-forming spaxels, fitting for radii between 0.5$\mathrm{R_e}$ and 2 $\mathrm{R_e}$. We perform error estimations slightly differently in this case: we re-fit gradients 100 times apiece with Gaussian noise added, with the noise level set by the model errors, before calculating dispersions of the re-fit gradients as before.

\begin{figure*}
\begin{center}
	\includegraphics[trim = 1.5cm 11.5cm 0cm 13cm,scale=1]{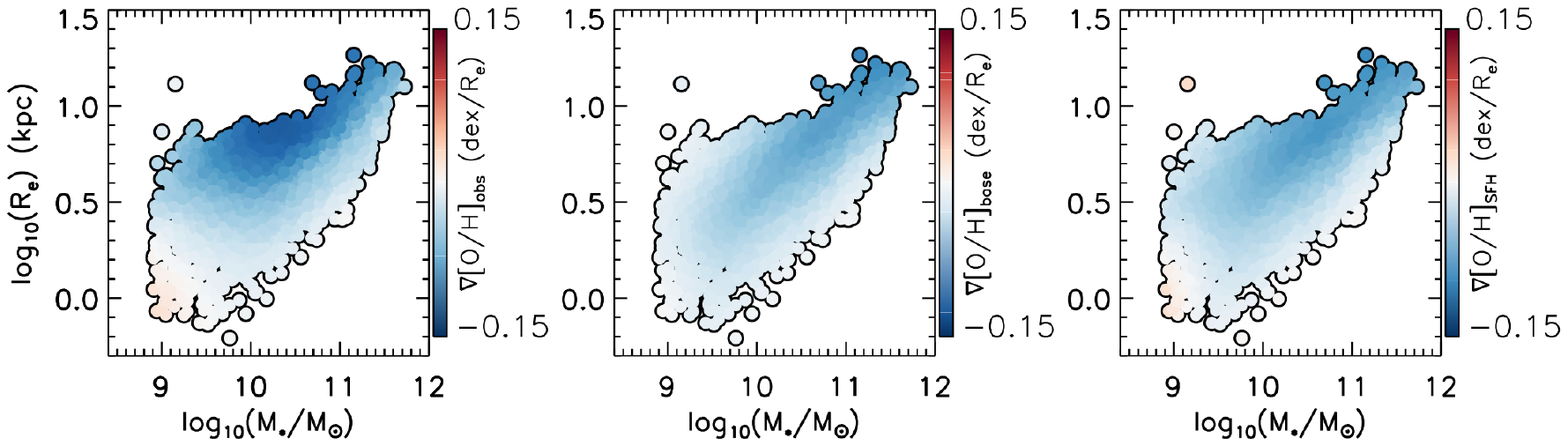}
	\caption{Effective radius, plotted against galaxy stellar mass, with data points coloured by the observed metallicity gradients (left; same as the bottom panel of \autoref{fig1}), the gradients predicted from the base models (middle) and the gradents predicted from the SFH models (right), with LOESS smoothing applied.}
	\label{fig6}
	\end{center}
\end{figure*}

We show in \autoref{fig6} the observed metallicity gradients and the model-predicted gradients across the mass-size plane, with LOESS smoothing applied in all cases. Immediately, we see that the predictions qualitatively reproduce two key features from the observations: the predicted gradients steepen with mass from the low-mass end, with a striking trend with mass--size position then emerging for the more massive galaxies such that larger galaxies display steeper gradients on average at a given mass. It is evident though that the most extended galaxies have predicted gradients somewhat flatter on average than what is observed, particularly at the low-mass end. 

For the base models, the median offset between the observed gradients and the base model gradients is -0.009 $\mathrm{dex/R_e}$ with a corresponding dispersion of 0.048 $\mathrm{dex/R_e}$. This dispersion is significantly higher than can be expected from measurement errors alone (typically below 0.02 $\mathrm{dex/R_e}$, with a median of 0.005 $\mathrm{dex/R_e}$) or from adding the observational measurement errors in quadrature with the model errors (which for the base models, produces a median error of 0.014 $\mathrm{dex/R_e}$). For the SFH models, we obtain a median gradient offset of -0.007 $\mathrm{dex/R_e}$ and a dispersion of 0.043 $\mathrm{dex/R_e}$. Thus, the SFH models achieve a reduced scatter in gradient offsets when compared to base models. At the same time, the dispersion in offsets remains higher than can be explained by measurement errors or by the scatter of local relations. 

\begin{figure*}
\begin{center}
	\includegraphics[trim = 1cm 16.5cm 1cm 7cm,scale=1.2,clip]{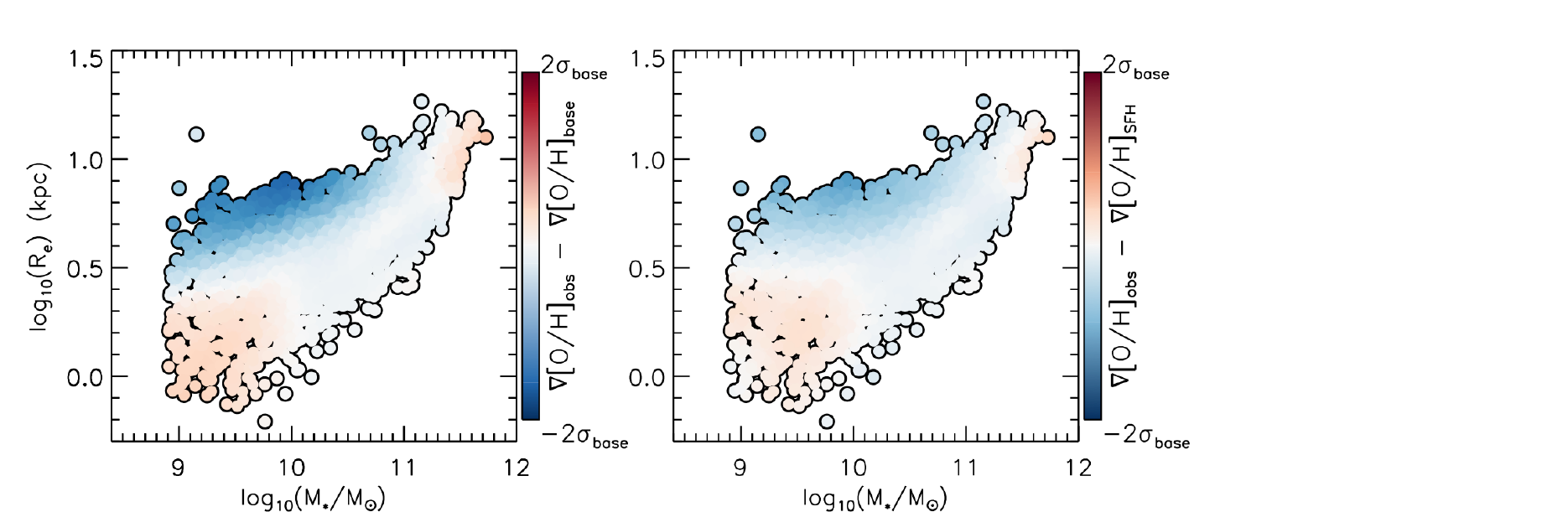}
	\caption{Effective radius, plotted against galaxy stellar mass, with data points coloured by the offsets between observed and model metallicity gradients. We have normalised the offsets by the dispersion of residuals from the base models, along with applying LOESS smoothing. We show results from the base models on the left, and from the SFH models on the right.}
	\label{fig9}
	\end{center}
\end{figure*}

We show in \autoref{fig9} the gradient residuals in mass--size space from the base models and SFH models, with LOESS smoothing applied; we have normalised these by the standard deviation of residuals between the observed and base model gradients. We find that the most extended galaxies typically possess steeper observed gradients than the models predict, particularly at low-to-intermediate masses. The average direction of the offsets reverses for small sizes and low masses, meanwhile. We find that the SFH models significantly reduce, but do not completely eliminate, this behavior. 

\subsection{Radial profiles of metallicities and residuals}

So far, we have demonstrated that local gas metallicity trends can qualitatively reproduce the metallicity gradient trend reported in B21: past stellar masses of around $10^{10}M_\odot$, larger galaxies have steeper predicted gradients on average at a given stellar mass. However, we also found quantitative differences even for our SFH models, with extended galaxies frequently displaying steeper observed gradients than the models predict. Thus, in this section we explore radial profiles of metallicities and residuals across the mass-size plane, for galaxies with well-predicted gradients and for galaxies with significant gradient offsets.

\begin{figure}
\begin{center}
	\includegraphics[trim = 1cm 16.5cm 1cm 7cm,scale=1.1,clip]{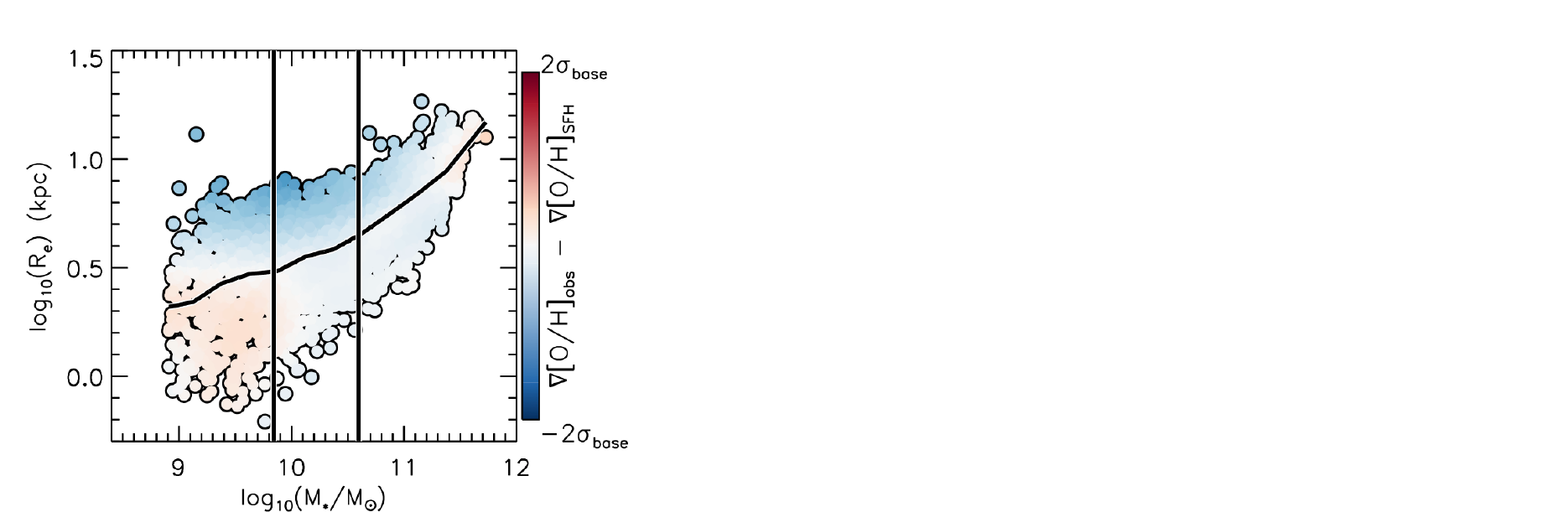}
	\caption{Same as the righthand window of \autoref{fig9}, but with the positions of our six mass-size bins overlaid.}
	\label{binfig}
	\end{center}
\end{figure}

To begin, we divide our sample into a series of six bins across the mass-size plane (\autoref{binfig}). We define three stellar mass regions selected to encompass 1/3 of the sample apiece -- $\mathrm{M_* \leq 10^{9.85}M_\odot}$, $\mathrm{10^{9.85}M_\odot < M_* \leq 10^{10.60}M_\odot}$ and $\mathrm{M_* > 10^{10.60}M_\odot}$ -- and term these regions ``low mass", ``mid mass" and ``high mass", respectively. We further split these regions according the median mass-size relation, which we calculate in bins of stellar mass: galaxies above the median relation are deemed ``extended", and galaxies at or below the relation are deemed  ``compact". By construction, these bins contain roughly equal numbers of objects, ranging from 345 galaxies (extended mid mass) to 362 galaxies (compact mid mass and extended high mass).

\begin{figure}
\begin{center}
	\includegraphics[trim = 8.5cm 8cm 0cm 3.5cm,scale=1,clip]{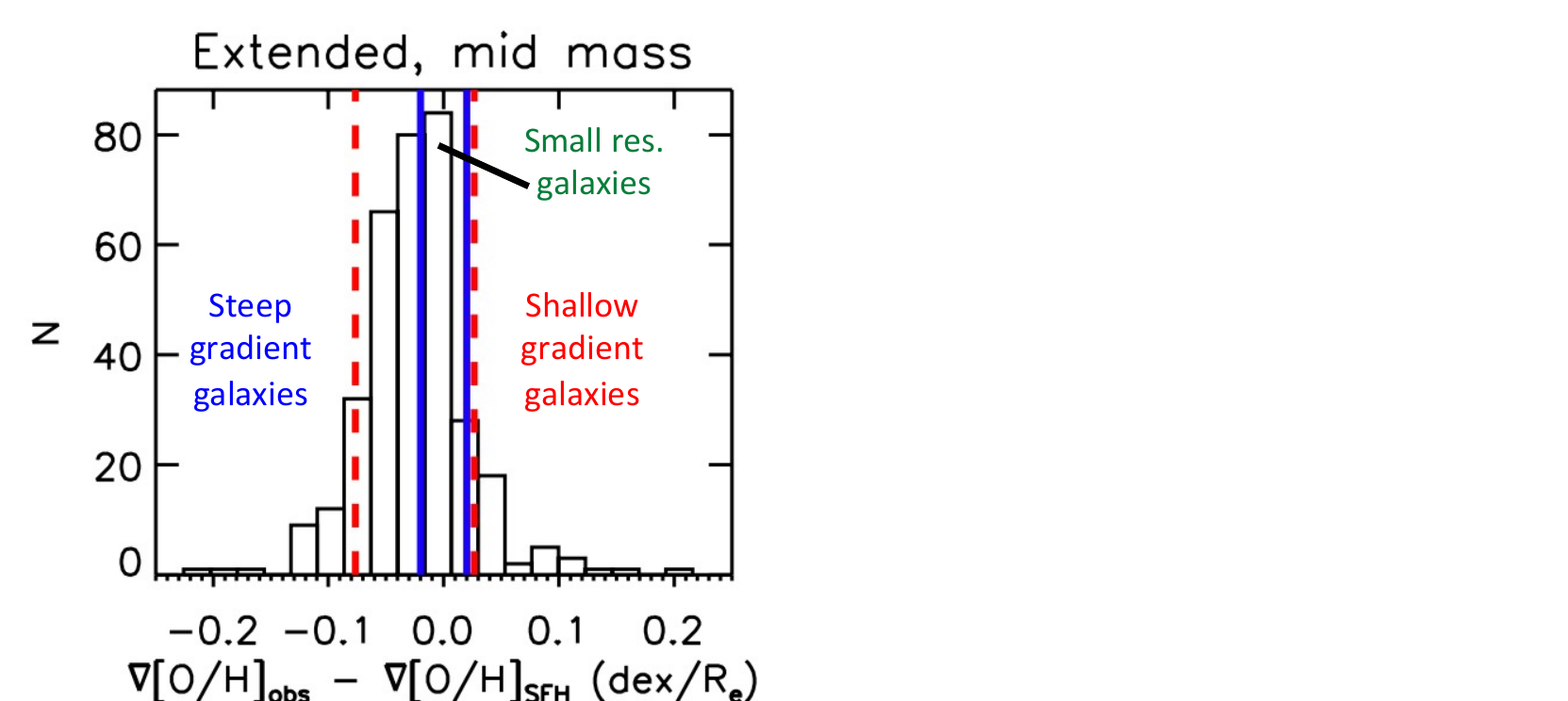}
	\caption{Demonstration of our subsample selection procedure, for the extended mid-mass bin. The red dashed lines indicate the 10th and 90th percentile regions; galaxies below the leftmost lines are selected as steep gradient galaxies, and galaxies above the rightmost line are selected as shallow gradient galaxies. Galaxies between the blue solid lines are selected as small residual galaxies.}
	\label{gradreshist}
	\end{center}
\end{figure}

We then select three subsamples from each of the six mass-size bins. For each bin, we select as ``steep gradient galaxies" those galaxies at or below the 10th percentile of the gradient offset distribution -- i.e., those galaxies with gradients far steeper than predicted by the model -- and we select as ``shallow gradient galaxies" those galaxies at or above the 90th percentile\footnote{These terms are a slight simplification, since some galaxies display positive metallicity gradients (see, for instance, \autoref{fig1}). However, these terms hold true for the vast majority of our sample, so we use them for simplicity's sake.} -- that is, galaxies with observed gradients that are much flatter than predicted. Finally, we select as ``small residual galaxies" those galaxies with gradient residuals no greater that $\pm$0.02~dex/$R_e$. We demonstrate this process for one mass-size bin in \autoref{gradreshist}. 

\begin{figure*}
\begin{center}
	\includegraphics[trim = 1.5cm 11.5cm 0cm 7cm,scale=0.95]{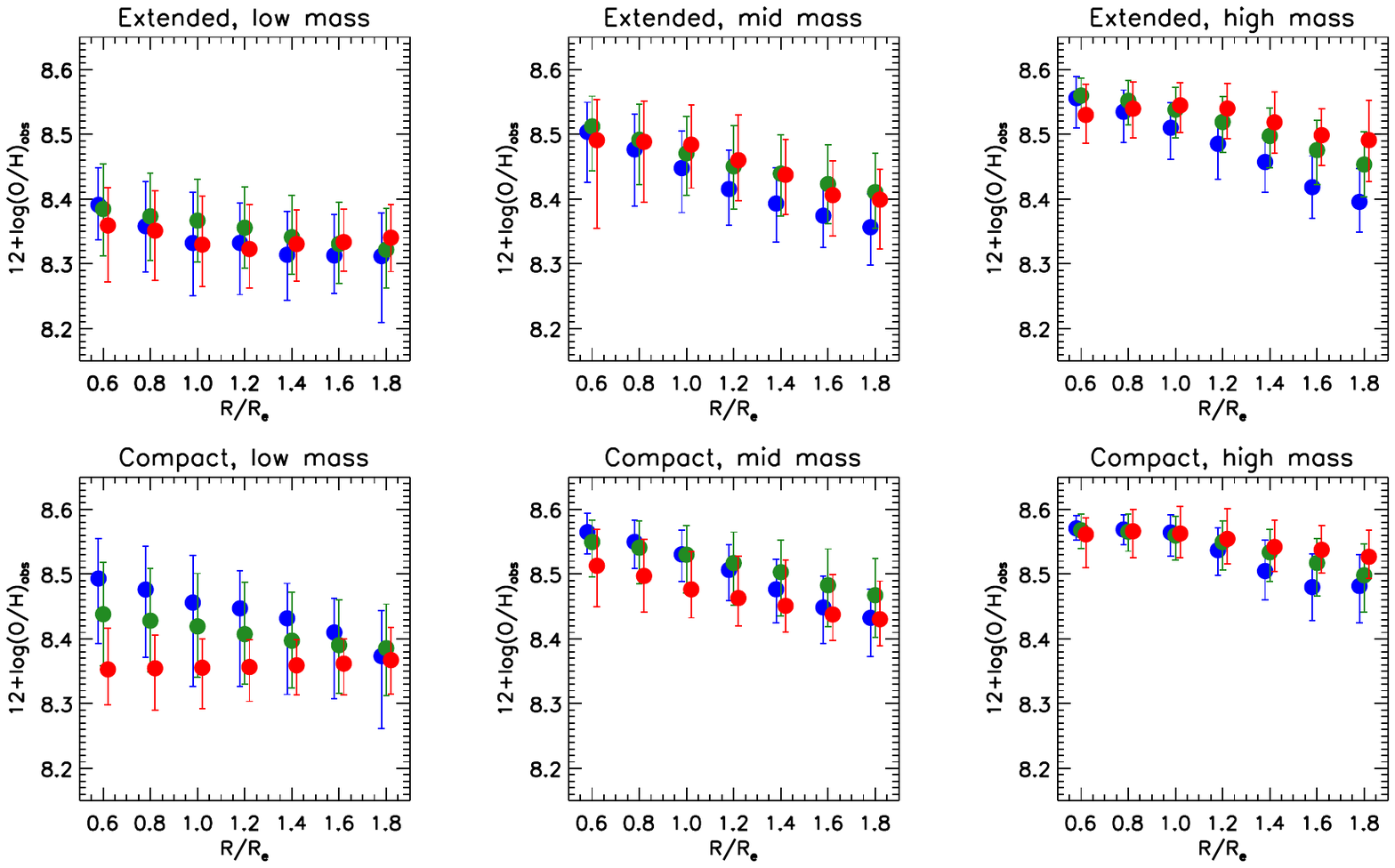}
	\caption{Radial profiles of observed metallicities, for steep gradient galaxies (blue points), small residual galaxies (green points) and shallow gradient galaxies (red points). Points show the median residuals within a given radial bin, with error bars encompassing the central 68\% of data points.}
	\label{zprof}
	\end{center}
\end{figure*}

We show the metallicity profiles of the subsamples in \autoref{zprof}. We calculate the profiles in radial bins of $\Delta r = 0.2~{\rm R_e}$ with boundaries between 0.5 $\mathrm{R_e}$ and 1.9 $\mathrm{R_e}$, calculating the medians along with the regions encompassing 68\% of data points. We continue to see different behaviour in different mass-size bins. For example, among compact low-mass galaxies, the ``steep gradient" galaxies have higher average metallicities at small radii when compared to the other subsamples. We see hints of similar behaviour in the mid mass bins. By contrast, shallow gradient galaxies in the high mass bins are typically more metal-rich at larger radii. In all cases, the metallicity profiles of steep gradient and shallow gradient galaxies are different on average from the profiles of small residual galaxies and from each other. In turn, we may speculate that the profiles of steep gradient and shallow gradient galaxies were shaped by processes that are \textit{not} fully captured by the local relations we study.

\begin{figure*}
\begin{center}
	\includegraphics[trim = 1.5cm 11.5cm 0cm 7cm,scale=0.95]{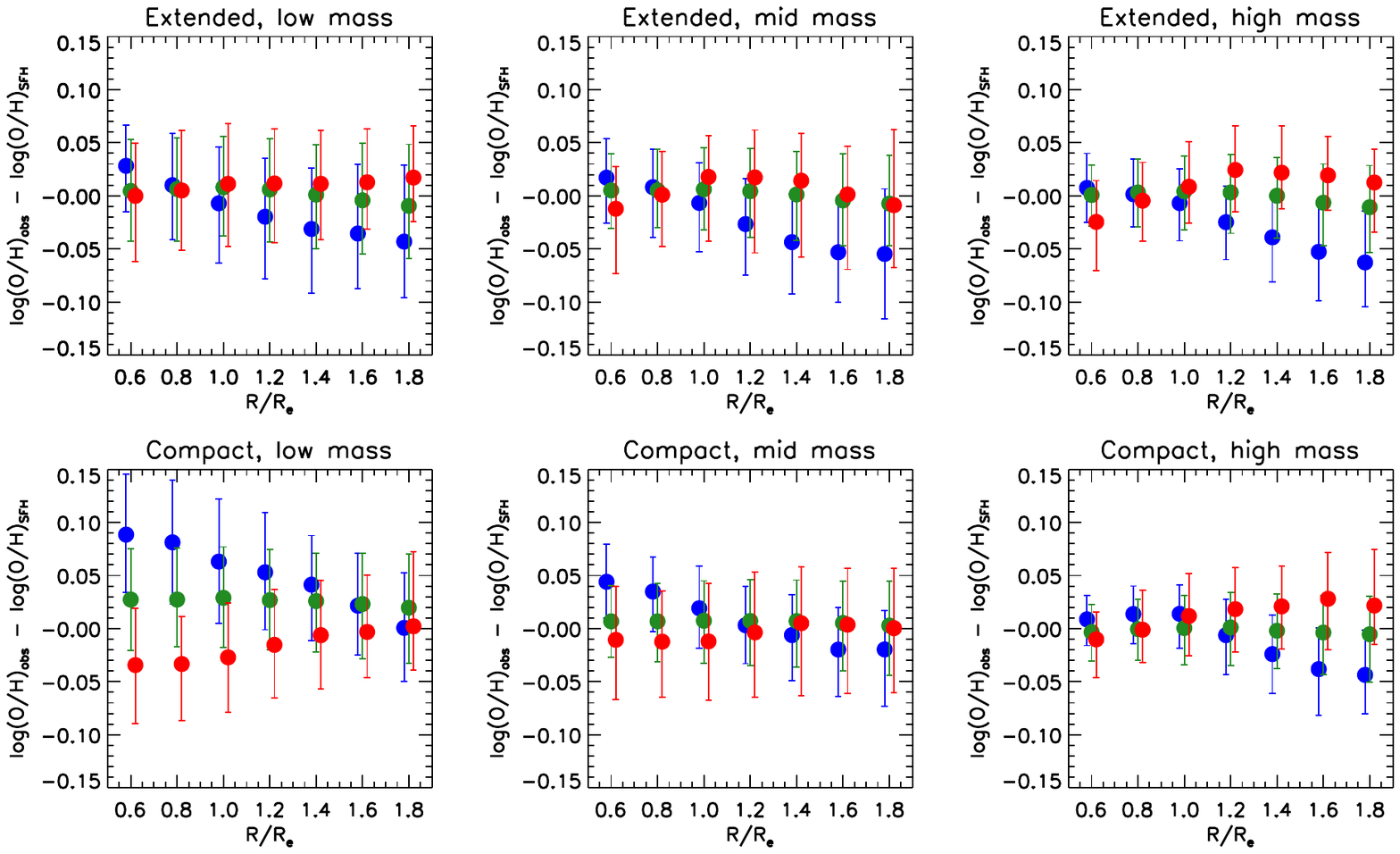}
	\caption{Radial profiles of residuals between observed and SFH model metallicities, for steep gradient galaxies (blue points), small residual galaxies (green points) and shallow gradient galaxies (red points). Points show the median residuals within a given radial bin, with error bars encompassing the central 68\% of data points.}
	\label{resprof}
	\end{center}
\end{figure*}

In \autoref{resprof}, we present the residual profiles between the observed and SFH model metallicities for these subsamples. We note similar slopes in the steep gradient galaxies across all mass-size bins, albeit with different normalisations. Steeper-than-predicted gradients are driven mainly by higher-than-predicted metallicities in the inner parts of compact low-mass and mid-mass galaxies and mainly by lower-than-predicted metallicities in galaxies' outer parts for the other four mass-size bins. 

\section{Discussion}\label{discussion}

\subsection{Empirical trends}

In B21, we demonstrated a striking trend in gas metallicity gradients as a combined function of galaxy mass and galaxy size. Here, we experimented with the use of local relations involving gas metallicity, with the aim of reproducing the observed mass-size gradient trend.

Individually, we found $\Sigma_*$ and then $M_*$ to present the strongest correlations with the observed metallicity (\autoref{modelsbase}). Given both this finding and the findings of previous works \citep[e.g.][]{gao2018}, we experimented with using the $M_*-\Sigma_*-\rm{O/H}$ relation to predict spaxel metallicities (our so-called base models). We found that this relation alone is largely sufficient to reproduce the gradients' behaviour in a qualitative sense: the gradients steepen with mass at low masses, and steepen with size at large masses, in agreement with observations.

We can therefore begin to understand the metallicity gradients across the mass-size plane by understanding the rMZR along with the mass-gradient connection, which itself has been reported before in MaNGA data \citep{belfiore2017,schaefer2020}. Intuitively, the rMZR can be understood as reflecting a connection between the density of a region and its star formation history \citep[e.g.,][]{bb2018}: denser galaxy regions formed the bulk of their stars earlier \citep[e.g.,][]{gonzalezdelgado2014} and so had more time to enrich. Given that the inner parts of galaxies are densest, metallicity gradients can also be understood in this manner. The relative flatness of low-mass galaxies' gradients is likely due to them being more sensitive to processes such gas mixing and wind recycling \citep[e.g.,][]{belfiore2017}, which will not be fully captured by the rMZR in isolation. 

The $M_*-\Sigma_*-O/H$ relation does not \textit{fully} reproduce the observed metallicity gradient behaviour in a quantitative sense. This is perhaps not surprising, as we find a number of residual metallicity trends in our spaxel sample (\autoref{modelsext}). For instance, we recover a local FMR from our spaxel sample, as reported previously for the O3N2 indicator in MaNGA data by \citet{teklu2020}; however, we are able to show here that this FMR is \textit{not} simply a projection of the residual $M_*$ dependence, and that it persists even once $M_*$ is taken into account.  It is evident that only a small minority of datapoints are significantly affected by the FMR as measured directly by $\Sigma_{SFR}$ and $\Sigma_*$; this is a possible reason why attempts at measuring the FMR on global scales with IFU samples -- which consist of relatively small samples of hundreds to thousands of galaxies -- have yielded non-detections \citep{bb2017,sanchez2013, sanchez2017a,sanchez2019a}. Use of EW(H$\alpha$) as a proxy may be a way forward for considering the global or local FMR in IFU galaxy sample samples, given its reduced susceptibility to error propagation when compared to the sSFR itself.

\citet{hwang2019} interpret the residual metallicity trends with $sSFR_{local}$, EW(H$\alpha$) and $\mathrm{D_n4000}$ as evidence of recent metal-poor gas inflows in affected regions, which serve to dilute metal content and to trigger renewed periods of star formation. Such gas could be obtained from gas-rich dwarfs or else from the intergalactic or circumgalactic medium, and could also result from enhanced radial inflows following interactions. A signature of this scenario is elevated N/O and enhanced SFR at fixed gas metallicity, as \citet{andrews2013} find for stacks of SDSS galaxies; such a signature is observed on $\sim$kpc scales, as reported by \citet{luo2021} from N/O measurements of MaNGA galaxies.

In regards to the \citet{hwang2019} scenario, we note with interest the trend between metallicity and $t_{LW}$ that we see for low-D4000 regions, along with the glut of low-D4000 datapoints with high $t_{LW}$ values evident in \autoref{dagelres}; a $t_{LW}$ bump is also evident in $t_{LW}$-EW(H$\alpha$) (\autoref{haewagelres}), albeit less prominently so. A possible explanation is that such regions recently rejuvenated their star-formation, which would be completely consistent with the arguments that \citet{hwang2019} put forth.

By correcting our base models for the metallicity residuals in $t_{LW}$-D4000 space, we constructed our so-called SFH models. We found that, compared to the base models, the SFH models yield improved agreement between observed and predicted metallicity gradients across the galaxy mass-size plane. This suggests that recent star formation histories help to drive mass-size gradient trends, and further supports a close connection between the metallicity of a star-forming region and its SFH.

However, the SFH models do not \textit{fully} eliminate the offsets we see between observed and predicted gradients, suggesting that a further explanation for the offsets is needed. Thus, it is worthwhile to more thoroughly consider the physical processes that shape the metallicity of a star-forming region.

\subsection{Effects of gas flows on metallicity gradients}

Observed gas metallicities are the products of chemical evolution, which can broadly be broken down into three key processes \citep[e.g.][]{ferreras2000}: gas infall, star formation (and subsequent ejection of metals), and metal outflow. We expect our ``SFH model" metallicities to indeed be sensitive to the star formation history of a given star-forming region. We also expect to be sensitive to variations in recent inflow rates, following the arguments of \citet{hwang2019}. Chemical evolution models support such an assertion, with time-varying inflow rates appearing sufficient to produce a negative SFR-metallicity correlation \citep{wang2021}.

However, our metallicity predictions are not necessarily sensitive to radial variations in outflow rates. Chemical evolution modelling has repeatedly pointed to outflow rates as a key factor for understanding gas metallicities \citep[e.g.][]{andrews2017}. Higher outflow rates are expected in regions of lower escape velocity \citep{bb2018}, and increased outflow rates in $\sim 10^9 M_\odot$ galaxies can explain their flattened metallicity gradients \citep{belfiore2019a}. Increased outflow rates at early times are also a potential explanation for discrepencies between gaseous and stellar metallicities, with a time-varying IMF providing an alternative explanation for this point \citep{lian2018,lian2018a}. Thus, variable present-day outflow rates -- which are not being captured by our metallicity predictions -- are a possible cause of remaining discrepencies in our predicted metallicity gradients.

Radial gas flows are another potential source of discrepencies between observed and predicted metallicity gradients. Recent simulations sugggest gas infall to be dominated by co-planar inflow events \citep[e.g.][]{trapp2022}.  However, observational evidence of significant radial gas inflows remains extremely limited: while \citet{schmidt2016} find mass inflow rates greater than star-formation rates for at least 5 of their 10 sample spiral galaxies, most such studies \citep{wong2004,trachternach2008,teodoro2021} do not generally detect significant radial flows in spirals. Furthermore, the expected effect of inward flows on metallicity profiles remains unclear, with steepening  and flattening \citep{kubryk2015, sharda2021} of gradients both suggested in different works.

To summarise: we expect our metallicity models to capture variations in recent star formation histories and in recent gas inflow rates, and we argue remaining gradient discrepencies to be due to physical processes that are \textit{not} well-captured by our models. Variations in recent outflow rates are possible explanation in this regard, as are the effects of radial gas flows.

\section{Summary \& Conclusions}\label{conclusion}

In B21, we demonstrated a striking trend in gas metallicity gradients across the mass-size plane: for galaxies at stellar masses of approximately $\mathrm{10^{10} \ M_\odot}$ and beyond, more extended galaxies display steeper gas metallicity gradients on average at a given stellar mass. This finding suggests that mass or size individually are not the best means to understand gas-phase metallicity gradients within a galaxy sample, and raises the question as to possible physical drivers of such behaviour.

Here, we set out to develop a physical interpretation of these observational results, by investigating the ability of local $\sim$kpc-scale trends to predict observed gas-metallicity trends. We constructed a set of metallicity predictions using galaxies' overall stellar masses along with their local stellar mass surface densities (which we deem our base models). We also experimented with corrections for various other residual trends connected to the SFH of star-forming regions.As part of these experiments, we noted a residual trend between metallicity and the light-weighted stellar age for low-D4000 regions, which to our knowledge has not been previously reported on the literature. We used this trend to construct a second set of model metallicities (which we refer to as SFH models). We argued that an age trend at low D4000 values could be explained by gas infall triggering renewed star formation in affected regions, which is entirely consistent with the arguments of \citet{hwang2019}).

Overall, we indeed reproduce the observational behaviour of galaxy gas metallicity gradients: at a given stellar mass beyond $\mathrm{10^{10} \  M_\odot}$, larger galaxies display steeper metallicity gradients (in units of dex/$\mathrm{R_e}$) on average. We also found the SFH models to yield improved gradient predictions over the base models. Thus, we argue that gas metallicity gradients can largely be understood in terms of local trends, which in turn can be understood as reflecting the connection between the formation history of a region and its observed metallicity. However, some average offsets can still be seen in the gradient predictions across the mass-size plane; we ascribe these to physical processes which are \textit{not} well-captured by the metallicity predictions, with variable outflow rates and radial gas flows providing potential explanations. 


A number of potential extensions to this work exist. In particular, it would be useful to test the use of escape velocity as an additional model parameter, give the suspected importance of metal outflows. Chemical evolution models of gas metallicity gradients across the mass-size plane would also be illuminating, in order to constrain the importance of outflows in setting the observed gradient trends. A comparison of metallicities with N/O abundance ratios in thus context would also be useful in light of the results of \citet{luo2021}, though the O/H calibrators employed in this work would not be suitable for such a study due to their use of the N2 indicator. Finally, a machine learning approach \citep[e.g.][]{bluck2019,bluck2020} would allow for a far more thorough investigation of how metallicity relates to other local parameters, and seems a logical way forward given the many-dimensional nature of the parameter space under study.

\section*{Acknowledgements}

N.B thanks Ricardo Schiavon for fruitful discussions during a visit to Liverpool John Moores University in  2020, which significantly influenced N.B's research direction towards work that would ultimately lead towards B21 and, hence, this present article.

We thank the anonymous referee for their thoughtful and constructive comments, which we feel have significantly improved the article.

 Funding for the Sloan Digital Sky Survey IV has been provided by the Alfred P. Sloan Foundation, the U.S. Department of Energy Office of Science, and the Participating Institutions. SDSS-IV acknowledges support and resources from the Center for High-Performance Computing at the University of Utah. The SDSS web site is \url{www.sdss.org}. J.B-B thanks IA-100420 (DGAPA-PAPIIT, UNAM) and CONACYT grant CF19-39578 support. RR thanks Conselho Nacional de Desenvolvimento Cient\'{i}fico e Tecnol\'ogico  ( CNPq, Proj. 311223/2020-6,  304927/2017-1 and 400352/2016-8), Funda\c{c}\~ao de amparo 'a pesquisa do Rio Grande do Sul (FAPERGS, Proj. 16/2551-0000251-7 and 19/1750-2), Coordena\c{c}\~ao de Aperfei\c{c}oamento de Pessoal de N\'{i}vel Superior (CAPES, Proj. 0001). RAR acknowledges financial support from Conselho Nacional de Desenvolvimento Cient\'ifico e Tecnol\'ogico (302280/2019-7).

SDSS-IV is managed by the Astrophysical Research Consortium for the Participating Institutions of the SDSS Collaboration including the Brazilian Participation Group, the Carnegie Institution for Science, Carnegie Mellon University, the Chilean Participation Group, the French Participation Group, Harvard-Smithsonian Center for Astrophysics, Instituto de Astrof\'isica de Canarias, The Johns Hopkins University, Kavli Institute for the Physics and Mathematics of the Universe (IPMU) / University of Tokyo, Lawrence Berkeley National Laboratory, Leibniz Institut f\"ur Astrophysik Potsdam (AIP),  Max-Planck-Institut f\"ur Astronomie (MPIA Heidelberg), Max-Planck-Institut f\"ur Astrophysik (MPA Garching), Max-Planck-Institut f\"ur Extraterrestrische Physik (MPE), National Astronomical Observatories of China, New Mexico State University, New York University, University of Notre Dame, Observat\'ario Nacional / MCTI, The Ohio State University, Pennsylvania State University, Shanghai Astronomical Observatory, United Kingdom Participation Group, Universidad Nacional Aut\'onoma de M\'exico, University of Arizona, University of Colorado Boulder, University of Oxford, University of Portsmouth, University of Utah, University of Virginia, University of Washington, University of Wisconsin, Vanderbilt University, and Yale University.

\section*{Data Availability}

All non-MaNGA data used here are publically available, as are all MaNGA data as of SDSS DR17.

\bibliographystyle{mnras}
\bibliography{bibliography}

\begin{appendix}

\section{SFH model parameter order}\label{ordertest_appendix}

For our SFH models, we started from our base model metallicities (in whcih metallicities are predicted using the $M_*-\Sigma_*-O/H$ relation) and then performed an additive correction according to the residuals in the $D4000$-$t_{LW}$ plane. Given the multi-dimensional nature of these models, it is worthwhile to consider if the order of applied parameters matters. To this end, we constructed two additional sets of models in an analogous manner to the SFH models. For the first set, we constructed ``base models in the $\Sigma_*-$D4000 plane and then corrected for residuals in the $M_*-t_{LW}$ plane; for the second set, we instead constructed ``base models" in the $M_*-$D4000 plane and then corrected for residuals in the $\Sigma_*-t_{LW}$ plane. As in the original SFH models, we required bins within a given parameter space to contain at least ten spaxels, with spaxels outside those bins not being assigned metallicities. We refer to these two new model sets as SFH-a models and SFH-b models respectively, to differentiate them from the original models.

\begin{figure}
\begin{center}
	\includegraphics[trim = 1.cm 11cm 1cm 7.5cm,scale=1.3]{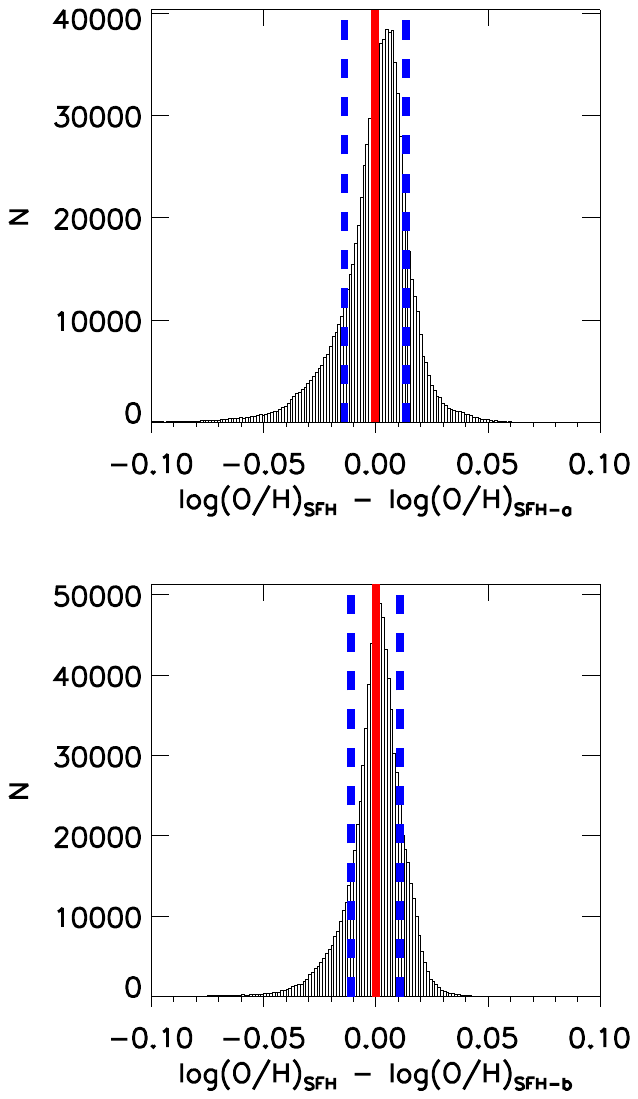} 
	\caption{Residuals between the SFH models and the SFH-a models (top) or the SFH-b models (bottom). The red solid line indicates the mean resdual, and the blue dashed lines the dispersion.}
	\label{otestfig}
	\end{center}
\end{figure}

To compare these two new models to the original SFH models, we selected all spaxels with assigned metallicities for all three model sets; this yielded a sample of 858065 spaxels. We then studied the scatter between the original SFH models with the other two, as shown in \autoref{otestfig}. We obtain median residuals close to zero, and we obtain residual dispersions significantly smaller than the data-model scatter. Thus, we can conclude that the order in which we apply parameters to our models has only a modest effect on the model metallicities. 

Two further compare the three models, we present in \autoref{ordertesttable} a comparison of model performances, assessiong the data-model residual dispersions along with the $\chi^2$ and BIC. We find the SFH models to yield the lowest dispersions and the lowest values of the BIC and $\chi^2$, indicating that the SFH models are superior to the other two in a statistical sense. Thus, the order of model parameters \textit{does} affect the model outputs, though the impact is indeed modest in practice.

\begin{table*}
\begin{center}
\begin{tabular}{c|c|c|c|c|c}
Model set & Model parameters & Residual dispersion & $\chi^2$ & k & BIC \\
\hline
\hline

SFH-a models & $\Sigma_*$, D4000, $M_*$,  $t_{LW}$ & 0.0454 dex & $5.62 \times 10^6$  &4773 &$5.69 \times 10^6$ \\
SFH-b models & $M_*$, D4000, $\Sigma_*$, $t_{LW}$ & 0.0466 dex & $5.13 \times 10^6$  &4526 &$5.19 \times 10^6$ \\
\textbf{SFH models} & $\mathbf{M_*}$, $\mathbf{\Sigma_*}$, \textbf{D4000}, $\mathbf{t_{LW}}$ & \textbf{0.0452 dex} & $\mathbf{4.97 \times 10^6}$  &\textbf{4267} &$\mathbf{5.03 \times 10^6}$ \\
\end{tabular}
\end{center}
\caption{Comparison of performance between the original SFH models and varients for which parameters are treated in different orders. We find the original SFH models to perform best in a statistical sense.}
\label{ordertesttable}
\end{table*}

\section{Results from the R2 calibrator}\label{p16appendix}

Over the course of this article, we have focused on results from a single gas metallicity calibrator (specifically, the O3N2 calibrator of M13). However, different calibrators can yield significantly different results, in terms of both gas metallicities and gas metallicity gradients \citep[e.g.,][]{kewley2008,belfiore2017,sanchez2017a,teimoorinia2021}. In addition, \citet{schaefer2020} show that gas metallicity gradients are vulnerable to biases from N/O variations when calculated with the O3N2 indicator. Thus, we briefly present results obtained from the R2 calibrator of \\citet[][hereafter PG16]{pilyugin2016} (their Equations 4 and 5), using the same final galaxy sample presented in the main paper text. As with the M13 O3N2 calibrator, the P16 R2 calibrator was derived from empirical fitting of observational data. The R2 calibrator is much less vulnerable to biases from N/O variations, however, as it employs [OII]$_{3737, 3729}$ in addition to [OIII], [NII] and Hydrogen lines. Since the R2 calibrator employes the full [OIII] and [NII] doublets, we assume a fixed 1/3 ratio between the dominant and sub-dominant [OIII] and [NII] components.

\begin{figure*}
\begin{center}
	\includegraphics[trim = 1.5cm 11cm 1cm 13cm,scale=0.9]{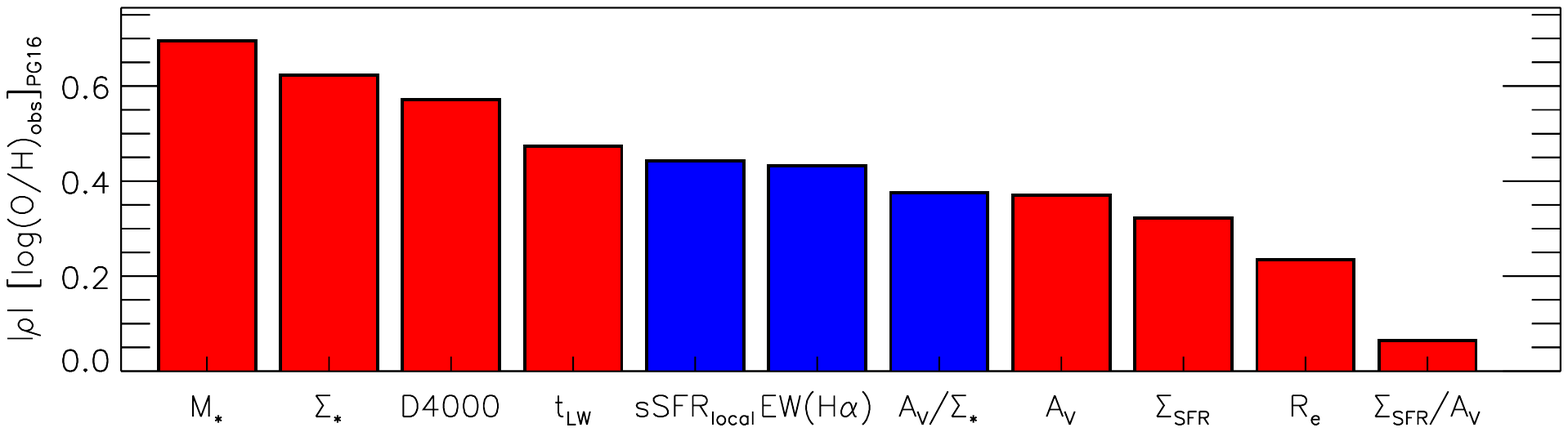} 
	\caption{Absolute values of the Spearman correlation coefficient between the R2-derived metallicity and various other parameters. Red bars indicate positive coefficients, and blue bars negative coefficients. We obtain $p \ll 0.01$ in all cases.}
	\label{obscoeff_pg16}
	\end{center}
\end{figure*}

\begin{figure}
\begin{center}
	\includegraphics[trim = 2cm 18cm 1cm 5cm,scale=0.95,clip]{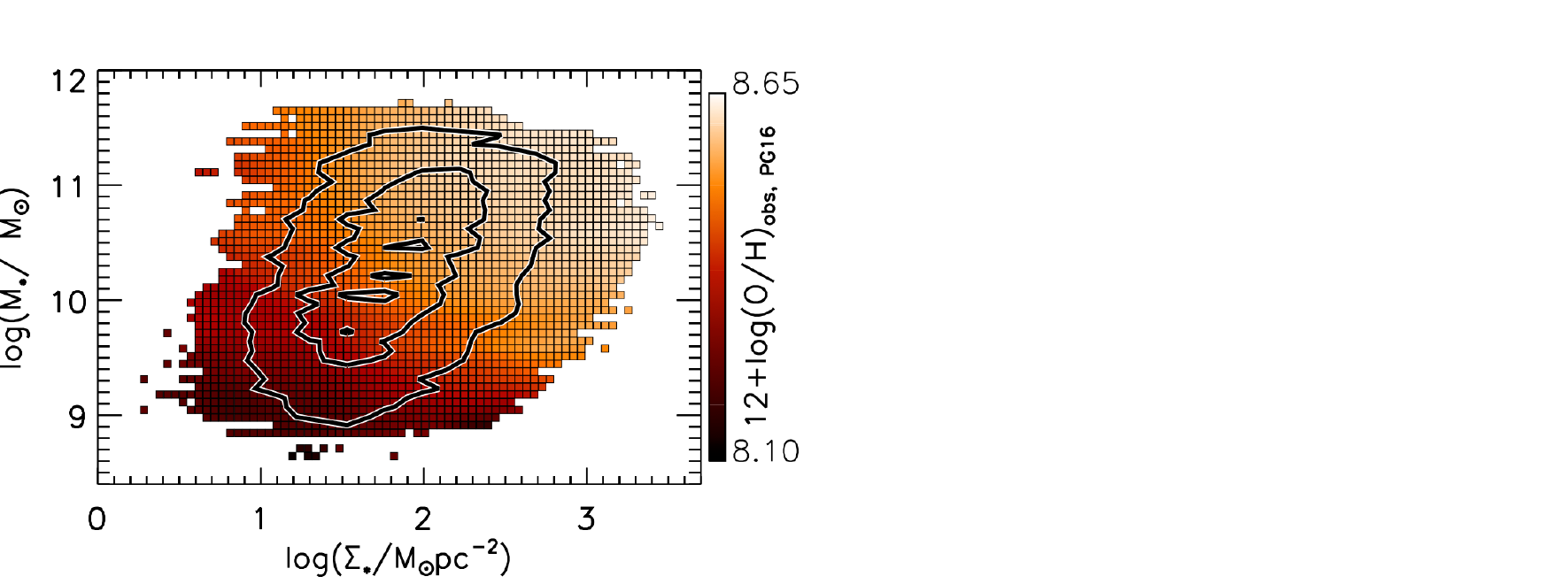} 
	\caption{Mean observed gas metallicity from the PG16 calibrator, as a function of $M_*$ and $\Sigma_*$, with weak smoothing applied. The contours encompass $\sim$10\%, $\sim$50\% and $\sim$90\% of sample galaxy spaxels.}
	\label{rhomfig_pg16}
	\end{center}
\end{figure}

We present in \autoref{obscoeff_pg16} the Spearman correlation coefficients between the R2-derived metallicity and all other considered parameters. As was found previously, we find $\Sigma_*$ and $M_*$ to yield the strongest individual trends, though we note higher scatter (lower coefficients) compared to the M13 case. We then construct a new set of ``base models" by computing the mean metallicity in bins of  $\Sigma_*$ and $M_*$, as demonstrated in \autoref{rhomfig_pg16}.

\begin{figure*}
\begin{center}
	\includegraphics[trim = 1.5cm 11.5cm 1cm 12cm,scale=1]{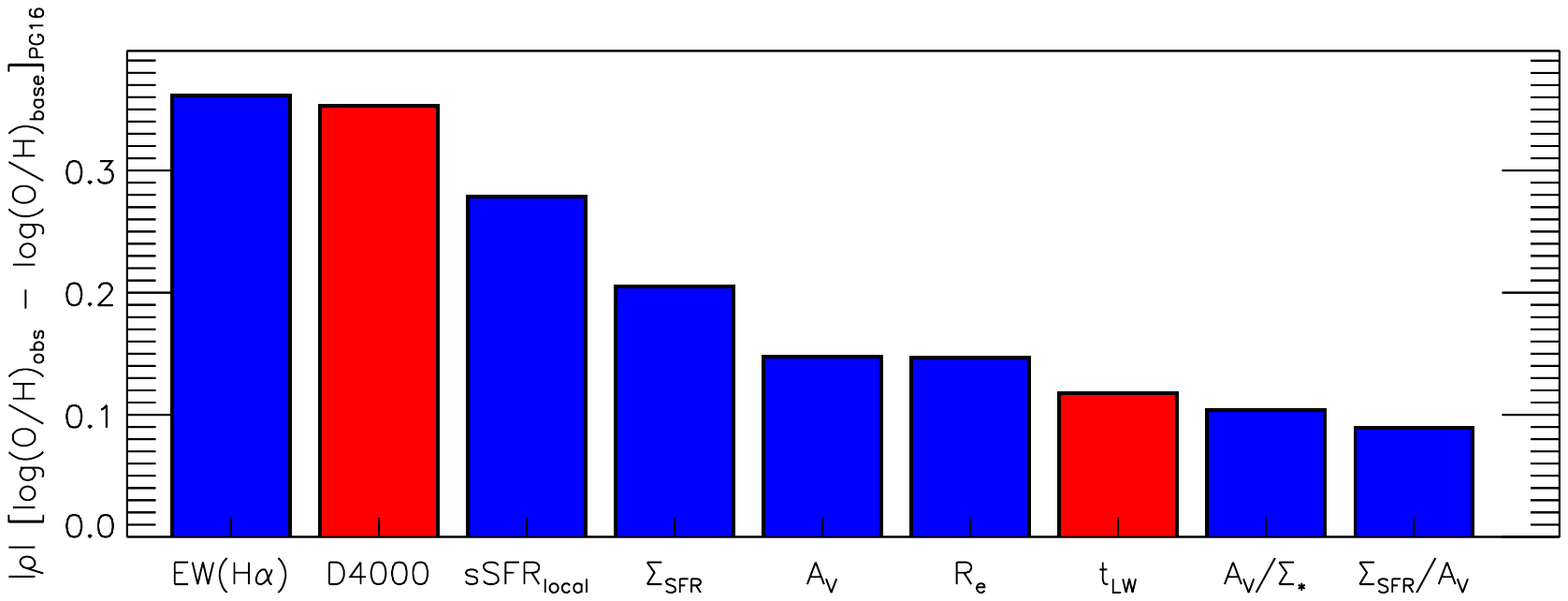} 
	\caption{Absolute values of the Spearman correlation coefficient between the gas-phase metallicity residuals and various other parameters. Red bars indicate positive coefficients, and blue bars negative coefficients. We obtain $p \ll 0.01$ in all cases.}
	\label{basecoeff_pg16}
	\end{center}
\end{figure*}

\begin{figure}
\begin{center}
	\includegraphics[trim = 4.cm 2.5cm 0cm 13cm,scale=0.9,clip]{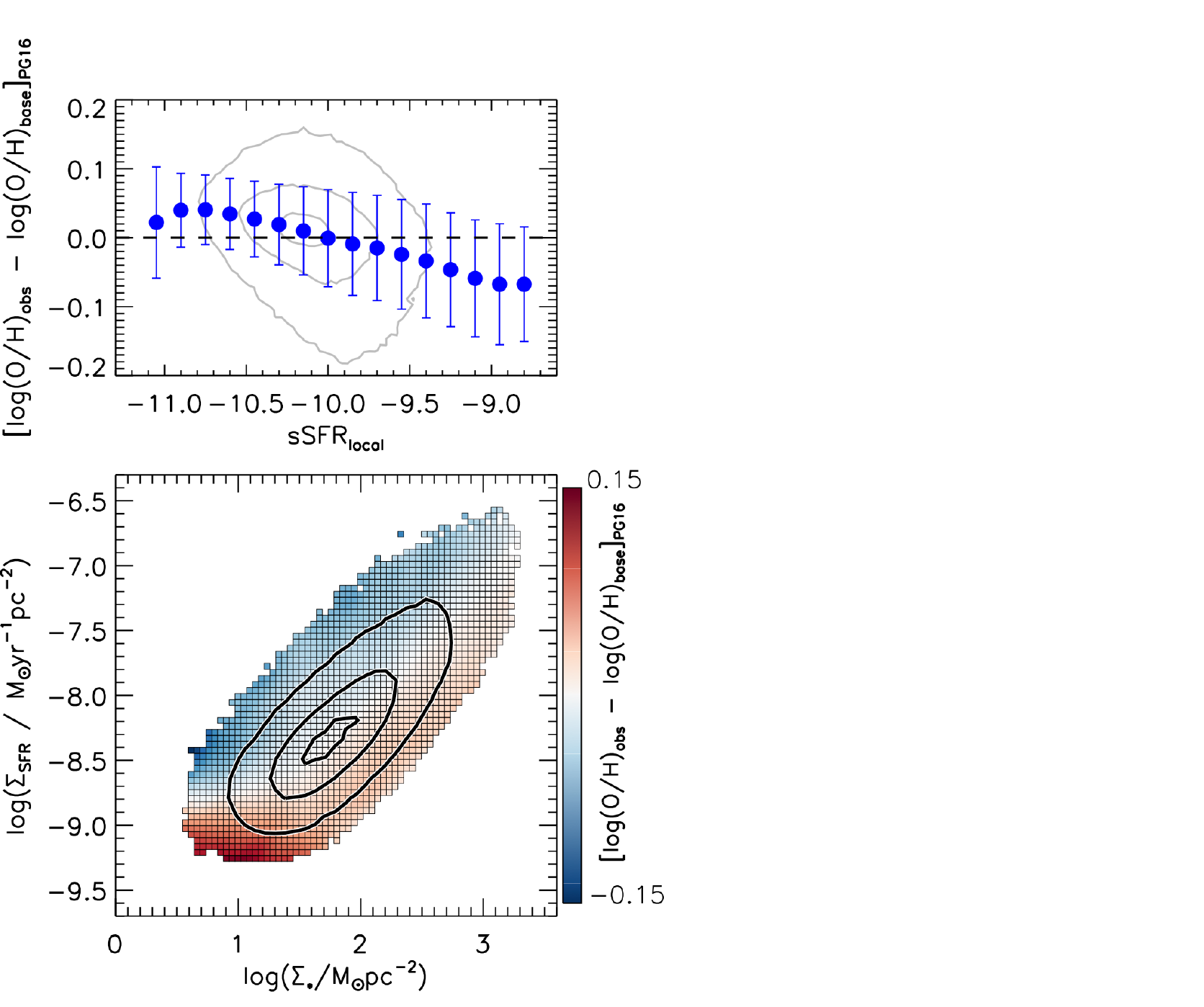} 
	\caption{Mean residuals between observed and base model metallicities with the PG16 calibrator employed, as a function of $sSFR_{local}$ (top) and as a combined function of $\Sigma_*$ and $\Sigma_{SFR}$ with weak smoothing applied (bottom). Contours encompass $\sim$10\%, $\sim$50\% and $\sim$90\% of sample galaxy spaxels.}
	\label{rhomsfrres_pg16}
	\end{center}
\end{figure}

From the new base models, we find the strongest residual metallcity correlations to be with D4000, EW(H$\alpha$) and $sSFR_{local}$ like before, though the correlations for all parameters but D4000 and EW(H$\alpha$) are somewhat stronger than was found from the M13 calibrator. We show $\rho$ values between the residuals and all parameters besides $M_*$ and $\Sigma_*$ in \autoref{basecoeff_pg16}. In \autoref{rhomsfrres_pg16}, meanwhile, we show the residuals as a function of $sSFR_{local}$ and as a combined function of $\Sigma_*$ and $M_*$; we see from these plots that the local FMR is detected in this dataset when the R2 calibrator is applied.

\begin{figure}
\begin{center}
	\includegraphics[trim = 0.6cm 11cm 1cm 10cm,scale=0.77,clip]{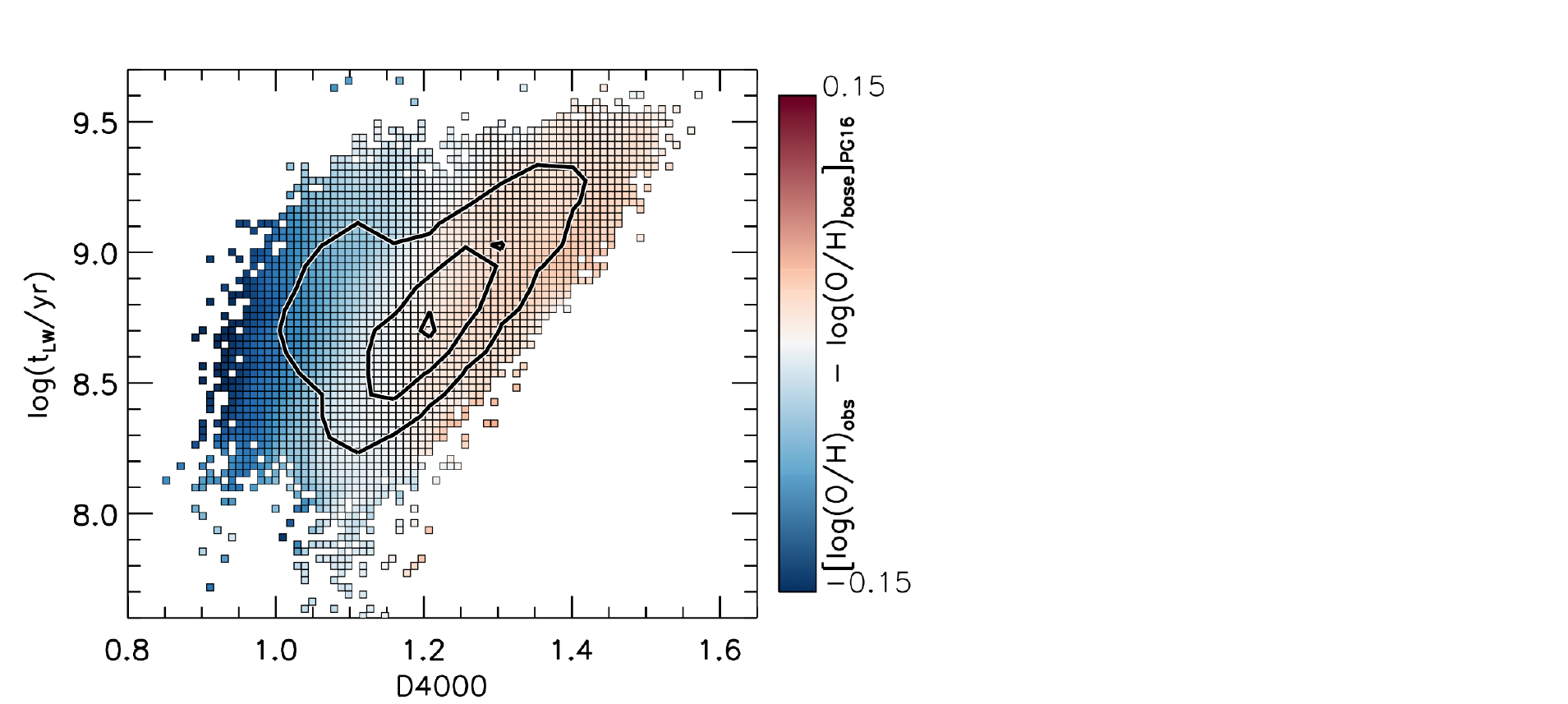}
	\caption{Mean residuals between observed and base model metallicities from the PG16 estimator, as a function of D4000 and \textbf{$t_{LW}$}. The contours encompass $\sim$10\%, $\sim$50\% and $\sim$90\% of sample galaxy spaxels.}
	\label{dagelres_pg16}
	\end{center}
\end{figure}

In \autoref{dagelres_pg16}, we plot the base model residuals as a combined function of D4000 and $t_{LW}$, from which we again detect a residual $t_{LW}$ trend at low values of D4000; by correcting for this trend, we produce a new set of ``SFH models" in the same manner as we did with the M13 metallicities previously.

From the galaxy sample we employed in the main paper text, we obtain a median spaxel metallicity offset of 0.007 dex for the base models and 0.002 dex for the SFH models. We obtain dispersions of 0.069 dex (base models) and 0.062 dex (SFH models).

\begin{figure*}
\begin{center}
	\includegraphics[trim = 1.5cm 11.5cm 0cm 13cm,scale=1]{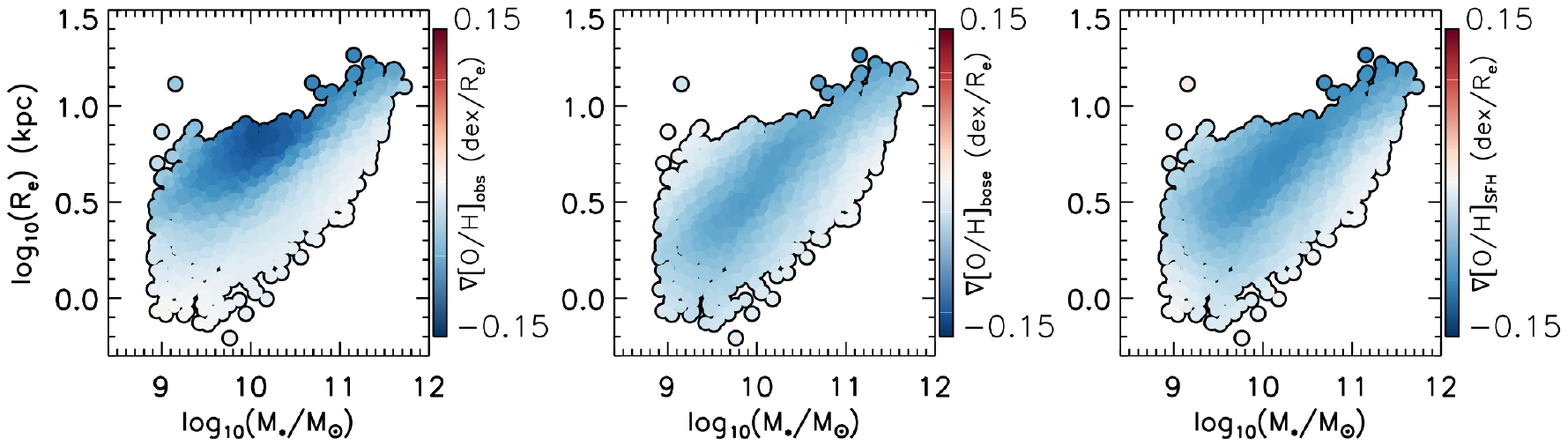}
	\caption{Effective radius, plotted against galaxy stellar mass, with data points coloured by the observed metallicity gradients (left; same as the bottom panel of \autoref{fig1}), the gradients predicted from the base models (middle) and the gradents predicted from the SFH models (right), with LOESS smoothing applied and with the R2 calibrator employed.}
	\label{fig6_pg16}
	\end{center}
\end{figure*}

\begin{figure*}
\begin{center}
	\includegraphics[trim = 1cm 16.5cm 1cm 7cm,scale=1.2,clip]{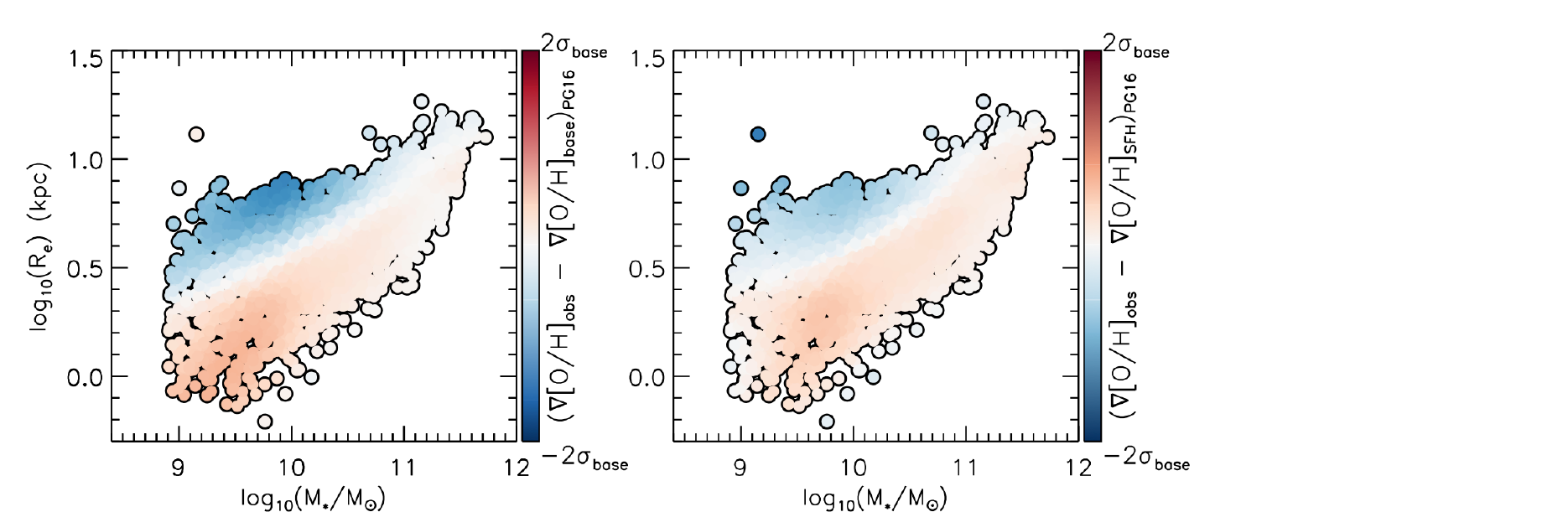}
	\caption{Effective radius, plotted against galaxy stellar mass, with data points coloured by the offsets between observed and model metallicity gradients in the case where the PG16 calibrator is applied. We have normalised the offsets by the dispersion for each set of models, along with applying LOESS smoothing. We show results from the base models in the left window, and we show SFH model results in the right window}
	\label{fig9_pg16}
	\end{center}
\end{figure*}

In \autoref{fig6_pg16}, we present the observed metallicity gradients from the PG16 calibrator along with the gradients predicted from the base and SFH models, using the same galaxy sample as for the M13 case. We present in \autoref{fig9_pg16}, meanwhile, the offsets between observed and predicted metallicity gradients for both model sets. The offsets are normalised by the dispersion in the base model offsets, with LOESS smoothing also applied. We calculate median offsets of 0.005 dex/$\mathrm{R_e}$ (base models) and 0.009 dex/$\mathrm{R_e}$ (SFH models); we calculate offset dispersions of of 0.057 dex/$\mathrm{R_e}$ (base models)  and 0.052 dex/$\mathrm{R_e}$ (SFH models).

From the above two figures, we obtain a picture near-identical to that which we obtained from the M13 calibrator. The models qualitatively reproduce the observed gradient behaviour across the mass-size plane. However, the gradient offsets themselves also trend across the mass-size plane, with the SFH models somewhat reducing but not eliminating this behaviour when compared to the base models. Thus, we may conclude that the key results of this paper are not unique to the M13 metallicity estimator.

\end{appendix}

\label{lastpage}
\end{document}